\documentclass[aps,nofootinbib,pre]{revtex4}
\usepackage{epsfig,graphics,amssymb,amsmath,subeqnarray,setspace,graphicx,amsthm,subfigure}

\def\b{\mathbf}\def\u{\underline}

\begin{document}

\title{The optimal elastic flagellum}
\author{Saverio E. Spagnolie}
\email{sespagnolie@ucsd.edu}
\author{Eric Lauga}
\email{elauga@ucsd.edu}
\affiliation{Department of Mechanical and Aerospace Engineering, University of California, San Diego, 9500 Gilman Drive, La Jolla CA 92093-0411, USA.}

\date{\today}
\begin{abstract}
Motile eukaryotic cells propel themselves in viscous fluids by passing waves of bending deformation down  their flagella. An infinitely long flagellum achieves a hydrodynamically optimal low-Reynolds number locomotion when the angle between its local tangent and the swimming direction remains constant along its length. Optimal flagella therefore adopt the shape of a helix in three dimensions (smooth) and that of a sawtooth in two dimensions (non-smooth). Physically, biological organisms (or engineered micro-swimmers) must expend internal energy in order to produce the waves of deformation responsible for the motion. Here we propose a physically-motivated derivation of the optimal flagellum shape. We determine analytically and numerically the shape of the flagellar wave  which leads to the fastest swimming  while minimizing an appropriately-defined energetic expenditure. Our novel approach is to define an energy which includes not only the work against the surrounding fluid, but also (1) the energy stored elastically in the bending of the flagellum, (2) the energy stored elastically in the internal sliding of the polymeric filaments which are responsible for the generation of the bending waves (microtubules), and (3) the viscous dissipation due to the presence of an internal fluid. This approach regularizes the optimal sawtooth shape for two-dimensional deformation at the expense of a small loss in hydrodynamic efficiency.  The optimal waveforms of finite-size flagella are shown to depend upon a competition between rotational motions and bending costs, and we observe a surprising bias towards half-integer wave-numbers. Their final hydrodynamic efficiencies are above 6\%, significantly larger than those of swimming cells, therefore indicating available room for further biological tuning.
\end{abstract}
\maketitle

\section{Introduction}

\indent The locomotive capabilities of microorganisms are intimately tied to the properties of the surrounding fluid medium \cite{lp09}. On scales relevant to most microorganisms, inertial effects are dominated by viscous dissipation; hence, the ejection of momentum into the fluid by the shedding of vortices, as observed in the locomotion of fish and birds, is not a viable means of propulsion for bacteria and spermatozoa. Instead, microorganisms have evolved to exploit hydrodynamic drag. Biological locomotion in this regime is the topic of a vast body of research, and we refer the reader to an excellent introduction by Purcell \cite{Purcell77}, and the classic texts by Lighthill \cite{Lighthill75} and Childress \cite{Childress81}. One of the most commonly observed means of microorganismic propulsion is the propagation of periodic waves down the length of a slender flagellum. Drag anisotropy in viscous flows, in combination with the time-irreversibility of uni-directional beating patterns, renders this locomotive form one of rather few relatively efficient means of hydrodynamic propulsion in viscous fluids. 

Due to its ubiquity in Nature, flagellar locomotion has long attracted the attention of biologists, mathematicians, and engineers alike. Continuous advances in imaging have revealed new details regarding the structure and kinematics of eukaryotic flagella \cite{Macnab76,bfb91,twb00}, but theoretical considerations of flagellar locomotion extend back to the seminal works of Taylor \cite{Taylor51}, Hancock \cite{Hancock53}, Gray \cite{Gray55}, and Lighthill \cite{Lighthill76}. In these works the authors have considered the hydrodynamics of slender body locomotion, developed a resistive force theory for the relationship between velocities and forces, and have deduced consequences regarding possible, and in some cases optimal geometries. Corrections to the simplified resistive force theory are found in a more detailed slender body theory \cite{Batchelor70,Cox70,Lighthill76,kr76,Johnson80}. The comparison of theory to experiments was furthered significantly in the 1970s in the works of Machin \cite{Machin58}, Higdon \cite{Higdon79}, and Brokaw \cite{Brokaw65, Brokaw70, Brokaw72}. An excellent review article on flagellar and ciliary propulsion from that era is provided by Brennen and Winet \cite{bw77}. More recently, attention has been paid to the relationship between internal structure and hydrodynamics. Camalet and J\"ulicher \cite{cj00} have shown that periodic bending and sliding of the microtubule structure of axonemal flagella can lead to wave generation and organism propulsion. Reidel-Kruse et al. \cite{rhhj07} have considered the coordination of dynein motors in beating spermatozoa, and have argued that the only theoretical motor coordination that fits their experimental data is interdoublet sliding. Other avenues of current active research include the swimming dynamics of bodies in non-Newtonian fluid environments, such as the propulsion of spermatozoa in the human female reproductive tract \cite{Chaudhury79,Sturges81,fkp98,Lauga07,fpw07,sg09}.

It is natural to ask about the optimality of the flagellar shapes exhibited by nature. Lighthill \cite{Lighthill75} included a response to this question by maximizing a hydrodynamic efficiency over the passage of periodic waves down the length of an infinitely long flagellum. He showed that the optimal flagellar shape was one for which the angle between the local tangent to the flagellum and the swimming direction was constant. In three dimensions, this leads to an optimal flagella in the shape of a rotating helix, a swimming mechanism frequently observed in Nature.  In contrast, in two dimensions the optimal shape is non-smooth and adopts a sawtooth form.  Other early work in this vein was performed by Pirroneau \& Katz \cite{pk74}, who considered the hydrodynamically optimal shape of finite slender swimmers, and noted an amplitude to wavelength relation for optimality in sawtoothed and small amplitude sinusoidal waveforms. The optimal shapes of finite sawtoothed, sinusoidal, and other curves that are amenable to analysis have also been studied by Silvester \& Holwill \cite{sh72}, Higdon \cite{Higdon79}, and Dresdner et al.\cite{dkb80}. More recently, Tam \cite{tamPhD} has shown numerically that the optimal slender swimmer in a Stokes flow does in fact pass periodic waves along its length, limiting to nearly the sawtoothed result of Lighthill.

In this paper we consider a physically-motivated approach to the  question of optimality in planar flagellar locomotion by explicitly taking into account the internal nature of the flagellum. We study changes to the hydrodynamically optimal but non-smooth shape of Lighthill when internal energetic costs are included. Specifically, we determine analytically and numerically the shape of the flagellum which, through the passage of a wave down its length, swims the fastest while minimizing a newly-defined swimming energy. This energy not only includes dissipation in the surrounding fluid, but also (1) elastic energy stored in the bending of the flagellum, (2) the Hookean energy stored in the relative silding of the  polymeric filaments (microtubules) which create the waves of deformation, and (3) the internal dissipation due to the presence of a  fluid inside the axoneme. In the case of infinite flagellum length, we show that this approach regularizes the non-smooth solution of Lighthill, at the expense of a small decrease in the hydrodynamic efficiency (from 8.5\% to 7.5\%). Finite-length swimmers also display smooth flagellar shapes, and we show that optimal shape is determined by a competition between rotational motions and bending costs with a surprising bias towards half-integer wave-numbers. The hydrodynamic efficiencies of the optimal finite-length flagella are above 6\%, significantly larger than those of biological cells (typically in the 1\% range), indicating available room for further biological morphological tuning.

The paper is structured as follows.  In \S\ref{kinematics} we introduce the notation for the swimming kinematics, as well as the new swimming energy measures we use in the paper. The case of an infinite-length flagellum is treated analytically in \S\ref{infinite} by a variational approach, and solved  numerically for finite bending and sliding costs. The results for finite-size flagella are presented in 
\S\ref{finite}. We conclude with a discussion of our results and their implications for the biophysics of motility in \S\ref{discussion}.

\section{Kinematics, fluid-body Interaction, and energetic costs}\label{kinematics}

\subsection{Kinematics}
We consider the passage of a periodic waveform down along an inextensible flagellum of length $L$ and radius $a$, which is confined to motion in the $x-z$ plane. The waveform is described by $\b{X}(s)=(X(s),Z(s))$, where $s\in[0,L]$ is the arc-length ($|\b{X}_s(s)|=1$). The waveform is chosen so that the body is initially oriented along the x-axis, with $\b{X}(0)=0$, and
\begin{gather}
X(s+\Lambda)=X(s)+\lambda,\,\,\,\,\,Z(s+\Lambda)=Z(s),
\end{gather}
where $\Lambda$ is the distance along the flagellum between wavelengths. $\lambda$ is the physical wavelength, so that $\alpha=\lambda/\Lambda<1$ is a contraction factor due to the
waviness of the flagellum \cite{Lighthill75}. We define $L=k \Lambda$, with $k$ the number of wavelengths along the body (not necessarily integral). 

\begin{figure}[ht]
\begin{center}
\includegraphics[width=3in]{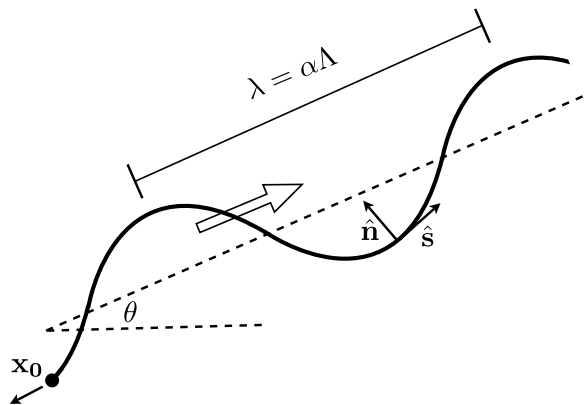}
\caption{Swimming flagellum with notation. A periodic waveform with physical wavelength $\lambda$ is passed from the head $\b{x_0}(t)$ down along the body centerline (to the right) at an angle $\theta(t)$ to the horizontal. The body moves opposite the direction of the wave in mean. Unit tangent and normal vectors are also indicated.}
\label{Figure1}
\end{center}
\end{figure}

The body motion is illustrated in Fig.~\ref{Figure1}. At time $t$, the waveform $(X(s),Z(s))$ is assumed to pass along the length of the body at an angle $\theta(t)$ to the x-axis. In a frame of reference moving with the traveling wave, the flagellum moves tangentially with uniform speed $c$ and a period $T^*=\lambda/c$. Defining $\b{x_0}(t)$ as the position of the head and $\b{r}(s,t)=\b{X}(s-c\,t)-\b{X}(-c\,t)$, the body centerline is written as
\begin{gather}
\b{x}(s,t)=\b{x_0}(t)+\u{\b{R}}\,\b{r}(s,t),
\end{gather}
where $\hat{i}=\cos(\theta)\hat{x}+\sin(\theta)\hat{z}$, and $\u{\b{R}}$ is the rotation operator,
\begin{gather}
\u{\b{R}}=\left(
            \begin{array}{cc}
              \cos\theta(t) & -\sin\theta(t) \\
              \sin\theta(t) & \cos\theta(t) \\
            \end{array}
          \right).
\end{gather}
Hence, the velocity of each point may be written in the lab frame as
\begin{gather}
\b{u}(s,t)=\b{\dot{x}_0}(t)+\u{\b{R}}\left(\b{r}_t+\dot{\theta}(t)\,\b{r}^\perp\right),
\label{Eqn:u}
\end{gather}
with $\b{X}^\perp = (-Z, X)$ and $\b{r}_t=-c\,\b{X}_s(s-c\,t)+c\,\b{X}_s(-c\,t)$. The unit tangent vector along the body (in the direction of increasing $s$) is denoted by $\b{\hat{s}}=\mathbf{\underline{R}(X}_s)$, and the normal vector is $\b{\hat{n}}=\b{\hat{s}}^\perp$.

\subsection{Fluid-body interactions}
The fluid-body interactions are modeled using the local resistive force theory of Gray \& Hancock (1955) \cite{gh55}. Resistive force theory relates the local fluid force per unit length and the local fluid/body velocity (equivalent by the assumption of a no-slip boundary condition). In the classical theory, the velocity of the slender body at a station $s$ is separated into tangential and normal components, as is the corresponding force per unit length $\b{f}(s,t)$:
\begin{gather}
\b{\hat{s}}\cdot \b{f}(s,t)=K_T\, \b{\hat{s}}\cdot\b{u},\,\,\,\,\,\,\b{\hat{n}}\cdot \b{f}(s,t)=K_N\, \b{\hat{n}}\cdot\b{u}.
\end{gather}
The force per unit length on the fluid, $\b{f}(s,t)$, is thus taken to be
\begin{equation}
\b{f}(s,t)=K_T \b{\hat{s}}\b{\hat{s}}^T\b{u}+K_N(I-\b{\hat{s}}\b{\hat{s}}^T)\b{u}=(K_T-K_N)\b{\hat{s}}(\b{\hat{s}}\cdot\b{u})+K_N\b{u}.
\label{Eqn:RFT}
\end{equation}
With no external forcing, the dynamics are thus set by conditions ensuring zero net force and zero net torque on the body at all times
\begin{gather}
\int_0^{L}\b{f}(s,t)\,ds=0,\,\,\,\,\,\, \int_0^{L}\left(\b{x}(s,t)-\b{x_0}(t)\right)\times \b{f}(s,t)\,ds=0\label{Eqn:ZeroForceTorquea}.
\end{gather}
These three equations are linear in the velocities $\b{\dot{x}_0}$ and $\dot{\theta}$, which are solved by a simple matrix inversion. The time-dependent body geometry determines uniquely the velocities at all times.

Corrections to the resistance coefficients, and to the local theory in general, are the subject of a number of studies \cite{gh55,Lighthill76,jb79}. For this study we fix the ratio $r_k=K_N/K_T=1/2$ with the acknowledgement that this ratio has been found in these other works to be dependent (though logarithmically) upon the ratio $a/\Lambda$ which we take to be very small. Non-local effects are potentially significant for the study of all but the thinnest of bodies; however, the use of slender body theory is complicated by the need for high resolution of the body shape near any regions of rapid geometric variation such as a kink. The slender body theory generates a system of Fredholm integral equations of the first kind which are in general susceptible to oscillatory behavior or slow convergence in their numerical solution, particularly when the immersed boundary has a sharp geometry \cite{Pozrikidis92}. This said, we note that Tam \cite{tamPhD} shows the near recovery of Lighthill's sawtoothed waveform solution using the full slender body theory. 

\subsection{Energetics}

\subsubsection{Dissipation}

The rate of mechanical work done by the body against the fluid, $\tilde{\Phi}^*(t)$,  is determined through an integration along the body centerline,
\begin{gather}
\tilde{\Phi}^*(t)=\int_0^L \b{f}(s,t)\cdot \b{u}(s,t)\,ds\,\,\,(\geq 0),
\end{gather}
and is seen to be non-negative due to the form of Eqn.~\ref{Eqn:RFT}. Averaging over one cycle, we define
\begin{gather}
\Phi^*=\langle \tilde{\Phi}^*(t) \rangle,
\end{gather}
where
\begin{gather}
\langle  \tilde{\Phi}^*(t) \rangle= \frac{1}{T^*}\int_0^{T^*}  \tilde{\Phi}^*(t) \,dt.
\end{gather}

In addition to performing work against the fluid, internal forces must also be exerted in order to create bending waves along the flagellum. In this paper we are considering these forces by explicitly taking into account the elastic nature of the flagellum, as well as the presence of internal dissipation. Three new measures of energy are therefore defined below. 

\subsubsection{Bending}
\begin{figure}[ht]
\begin{center}
\includegraphics[width=5.5in]{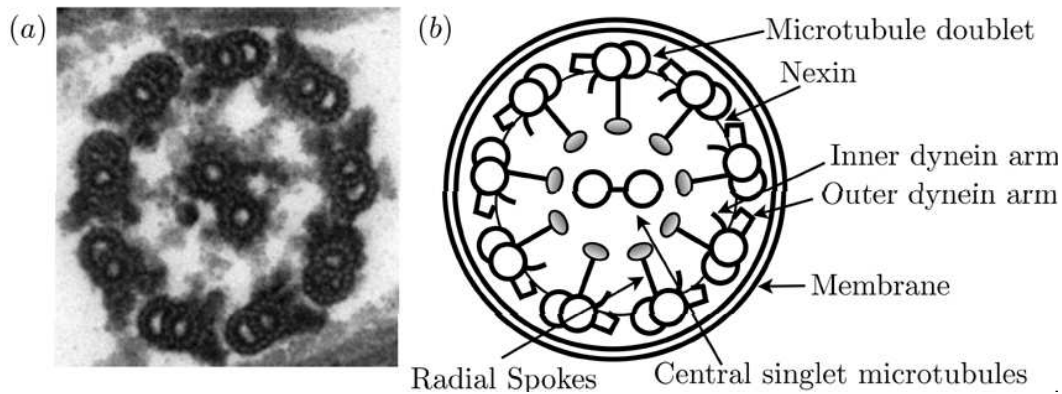}
\caption{The structure of the flagellar axoneme is seen in a (a) TEM image (Rippel Electron Microscope Facility, Dartmouth College) and (b) a cross-sectional diagram of a \textit{Chlamydomonas} flagellum.}
\label{Figure2}
\end{center}
\end{figure}

Figure~\ref{Figure2} shows a TEM image and a cross-sectional diagram of a typical Eukaryotic flagellum, in this case that of the organism \textit{Chlamydomonas}. The internal structure of a Eukaryotic flagellum, known as the axoneme, is usually composed of nine microtubule doublets which encircle a central microtubule pair (though other numbers and modifications of this basic pattern have been observed) \cite{bw77}. Dynein molecular motors act to generate shear forces that cause sliding between the outer doublet microtubules, and consequently the macroscopic passage of waves along the flagellar length \cite{Brokaw89}. Nexin proteins are elastic links that act to keep the outer microtubule doublets well spaced.

We model the elastic energy stored in the bending of the axoneme, $\mathcal{E}^*_{Bending}(t)$, as a function of the flagellum's effective Young's Modulus $E$, its second moment of inertia $I$, and the local flagellum curvature $\kappa(s,t)$,
\begin{gather}\label{bending}
\mathcal{E}^*_{Bending}(t)=E I\int_0^L \kappa^2(s,t)\,ds,
\end{gather}
(see \cite{ll86}). Kink instabilities have been shown to form when soft elastic cylinders are bent beyond a critical radius of curvature  \cite{gd07};  however, we take such defects to be negligible given the assumption of the vanishingly small aspect ratios considered here. A bending power associated with Eqn.~\eqref{bending} is defined as
\begin{gather}
\mathcal{P}^*_{Bending}=\langle\mathcal{E}^*_{Bending}(t)\rangle/T^*,
\end{gather}
and represents the time-averaged elastic energy stored in the flagellum per unit period of the wave.

\subsubsection{Elastic sliding}

\begin{figure}[ht]
\begin{center}
\includegraphics[width=2.75in]{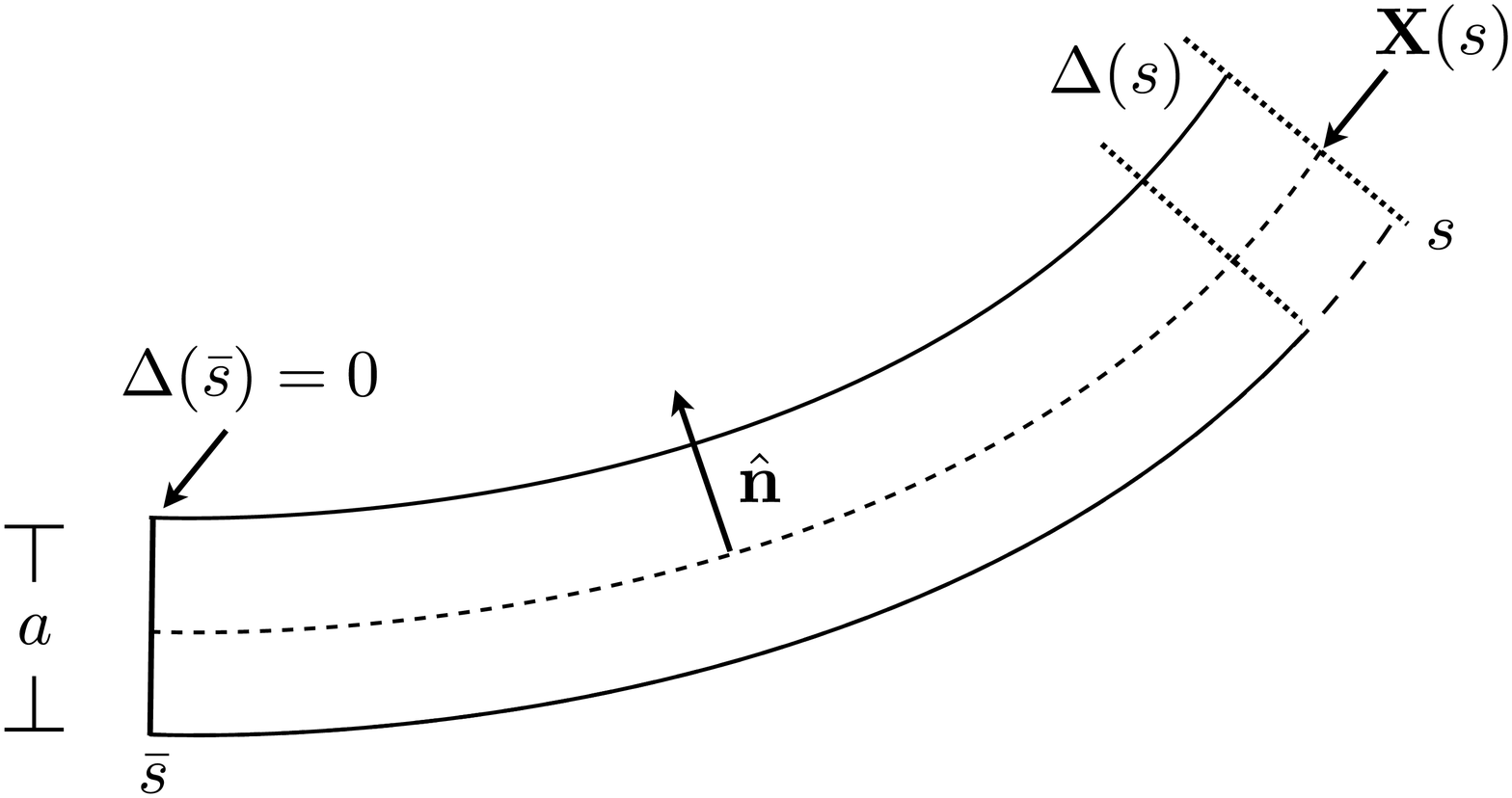}
\caption{Two-dimensional analogue of the axoneme shown in Fig.~\ref{Figure2}. A local sliding displacement $\Delta$ is defined as the difference between the arc-lengths of the top and bottom curves $\mathbf{X}(s)\pm (a/2)\hat{\mathbf{n}}(s)$. Here we have set zero relative displacement at the leftmost edge ($\Delta(\bar{s})=0$).}
\label{Figure3}
\end{center}
\end{figure}

The relative sliding between the microtubule doublets (Fig.~\ref{Figure2}) is understood to account for the generation of bending moments and large scale undulations \cite{cj00,rhhj07,Brokaw94}. In the study of planar waves, a common abstraction of the internal sliding is to consider a ``two-dimensional axoneme'' as illustrated in Fig.~\ref{Figure3} (following Camalet \& J\"ulicher \cite{cj00}). A rigorous connection between the two-dimensional consideration above and the full three-dimensional axoneme is presented in \cite{HilfingerPhD}, and bending moment propagation in flagella by such sliding action is considered in \cite{Brokaw71Sliding}.

To capture an energetic cost due to a material shear of this nature we define a sliding energy per wavelength under the assumption of a Hookean internal response. Based on the two-dimensional structure described above, we define a local sliding displacement $\Delta(s,t)$ as a difference between the arc-lengths of the top and bottom curves (as illustrated in Fig.~\ref{Figure3}),
\begin{align}
\Delta(s,t)&=\Delta(\bar{s},t)+\int_{\bar{s}}^s \,\left(\Big|\partial_s \left(\b{X}+\frac{a}{2} \b{\hat{n}} \right)\Big|-\Big|\partial_s \left(\b{X}-\frac{a}{2}\b{\hat{n}}\right)\Big|\right)\,ds'\\
&=\Delta(\bar{s},t)+a\int_{\bar{s}}^s\kappa(s',t)\,ds'\\
&=\Delta(\bar{s},t)+a\left(\psi(s,t)-\psi(\bar{s},t)\right).
\end{align}
Here we have defined the local tangent angle $\psi(s,t)$, where $X_s=\cos(\psi)$ and $Z_s=\sin(\psi)$, and $\psi_s=\kappa$. A complete description of the sliding distribution requires initial specification of the relative distance between top and bottom curves at a single reference point $\Delta(s=\bar{s},t=0)$. Subsequently for $t>0$ the sliding dynamics are set by the time-dependent geometry as we will show. The relative sliding displacement changes with the local shears generated by bending moments, and the behavior at $s=\bar{s}$ may be written in terms of a jump in upper and lower curve velocities,
\begin{gather}
\Delta_t(\bar{s},t)=[\b{\hat{s}}\cdot\b{u}]=a\,\b{\hat{s}}\cdot \b{\hat{n}}_t=-(a\,c)\,\b{\hat{s}}\cdot \b{\hat{n}}_s=-(a\,c)\,\psi_s(\bar{s},t),\label{Eq:Deltat}
\end{gather}
where $[f]=f^{\text{top}}-f^{\text{bottom}}$. Hence, the sliding displacement may be written as 
\begin{gather}
\Delta(s,t)=\Delta(\bar{s},0)-(a\,c)\,\int_0^t\,\psi_s(\bar{s},\tau)\,d\tau+a\left(\psi(s,t)-\psi(\bar{s},t)\right).
\end{gather}
We exploit the travelling wave structure, $\psi_s=-\frac{1}{c}\psi_t$, and integrate the above to give
\begin{gather}
\Delta(s,t)=\left(\Delta(\bar{s},0)-a\,\psi(\bar{s},0)\right)+a\,\psi(s,t)\\
=a\left(\,c_\Delta+\psi(s,t)\right).
\end{gather}

Reidel-Kruse et al. \cite{rhhj07} set $\bar{s}=0$ and couple the sliding displacement at the flagellar base, $\Delta(\bar{s}=0,t)$, to the internal sliding dynamics along the body length. A sliding energy is defined through a shear modulus $G$ as
\begin{gather}
\mathcal{E}^*_{Sliding}(t)=G \int_0^L \Delta^2\,ds,
\end{gather}
and we define an associated sliding power as the amount of sliding energy stored per unit period of the wave, 
\begin{gather}
\mathcal{P}^*_{Sliding}=\langle \mathcal{E}^*_{Sliding}(t) \rangle/T^*.
\end{gather}

\begin{figure}[ht]
\begin{center}
\includegraphics[width=2.5in]{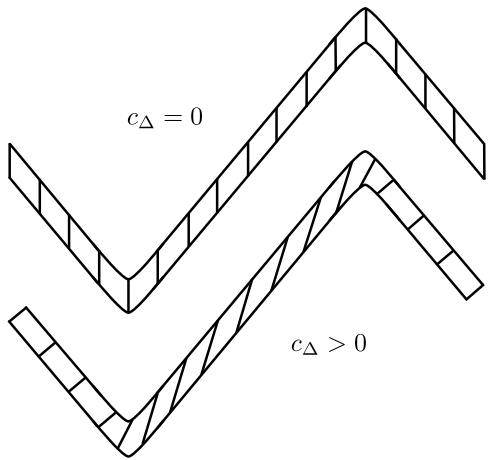}
\caption{Two arrangements of sliding displacement density are illustrated. The first corresponds to zero mean sliding displacement, $(a\,c_\Delta)=0$. The second corresponds to a positive mean sliding displacement, $(a\,c_\Delta)>0$, and hence a non-zero net internal moment.}
\label{Figure4}
\end{center}
\end{figure}

For periodic waveforms, $(a\,c_\Delta)$ is equivalent to the period-averaged sliding displacement, and we refer to $c_\Delta$ as a dimensionless base sliding. Having assumed a Hookean elastic response to sliding, $c_\Delta \neq 0$ therefore corresponds to a non-zero net internal moment. This moment would act, absent any other internal forces, to drive the flagellum towards the $c_\Delta=0$ state, or
\begin{gather}
\Delta(\bar{s},0)=a\,\psi(\bar{s},0),
\end{gather}
with the sliding displacement precisely equivalent to the tangent angle multiplied by the body thickness. This special case is illustrated in Fig.~\ref{Figure4}, along with an illustration of a body with non-zero base sliding, $c_\Delta>0$. For a given body shape, the second corresponds to a state with a larger internal energy. Absent external forces there can would also be a net body rotation in an energy minimizing response to $c_\Delta\neq 0$. 

\subsubsection{Rate of sliding}

The locomotive properties of many organisms may also depend upon the dynamics of a fluid internal to the body.  Hence, we consider a third cost of locomotion, that of internal dissipation due to the sliding of an upper and lower boundary as described above,
\begin{gather}
\tilde{\Phi}^*_{Internal}(t)=2\mu_I\int_0^L \left(\frac{\Delta_t}{a}\right)^2\,ds=2\mu_I\,c^2\,\int_0^L \kappa^2 \,ds,
\end{gather}
where $\mu_I$ is the viscosity of the internal fluid and we have used Eqn.~\ref{Eq:Deltat}. As a result, the optimal waveforms determined by including an elastic bending cost will also capture the effects on the optimal shape determined in the presence of an internal dissipation cost. This is at first glance a puzzling result; a shear of two surfaces certainly need not require a curvature. However, given the assumption of a periodic waveform and the associated velocities, the shears generated are in fact due only to the bending of the filament. Without loss of generality, we therefore consider below only elastic bending and sliding costs.

\subsubsection{Efficiency}
The system is made dimensionless by scaling velocities on the speed $c$, lengths on $L$ ($\Lambda$) for finite-length (infinite-length) bodies, and time on the ratio of length to velocity scale. Geometrical variables, velocities, and forces are henceforth understood to be dimensionless, and the period of motion is denoted by $T$. The rate of mechanical work, and bending and sliding powers for finite bodies are written as
\begin{gather}
\Phi=\Big\langle\int_0^1 \b{f}\cdot \b{u}\,ds\Big\rangle,\,\,\,\,\,\,\mathcal{P}_{Bending}=\gamma_{_B}\Big\langle\int_0^1 \kappa^2 \,ds\Big\rangle,\,\,\,\,\,\,\mathcal{P}_{Sliding}=\gamma_{_S}\Big\langle\int_0^1 \Delta^2 \,ds\Big\rangle,
\end{gather}
where $\gamma_{_B}$ and $\gamma_{_S}$ are dimensionless and may be inferred from the definitions above. Finally, we define a swimming efficiency. A common measure of hydrodynamic efficiency, $\eta_H$, for low Reynolds number swimming is the ratio of the rate of work required to drag the straightened flagellum through the fluid to the rate of work done to propel the undulating body at the same velocity,
\begin{gather}
\eta_H=\frac{r_k U^2}{\Phi},
\end{gather}
where $U=|\langle \b{\dot{x}_0}(t)\rangle|$ is the (dimensionless) mean swimming speed.

In this paper we define a generalized swimming efficiency, $\eta$, by including as well the rates of energetic expenditure due to internal bending (or internal dissipation) and sliding,
\begin{equation}
\eta= \frac{r_k U^2} {(1-A_B)(1-A_S)\Phi+(A_B/\gamma_{_B})\mathcal{P}_{Bending}+(A_S/\gamma_{_S})\mathcal{P}_{Sliding}}.
\label{E:eta}
\end{equation}
$A_B\in[0,1]$ and $A_S\in[0,1]$ are dimensionless numbers which allow for variation of the relative importance of bending and sliding energetic costs. The optimal swimmer is henceforth defined as that waveform $\b{X}(s)$ which maximizes the efficiency $\eta$. We note that an alternative measure of efficiency has been defined in Ref.~\cite{Childress81}; however, in order to compare our work most directly to the classical result of Lighthill and other recent works we have chosen the measure above.

\section{Bodies of Infinite-Length}
\label{infinite}

\subsection{A classical result by a variational approach}

We begin by showing that a variational approach yields the classic result due to Lighthill \cite{Lighthill75}. When the body is infinitely long ($L\rightarrow \infty$), it is useful to decompose the body velocity into the tangential motion along the waveform and a swimming velocity, $\tilde{\b{U}}=\langle \tilde{U},0\rangle$ (as in \cite{Lighthill75}) which we achieve by defining
\begin{gather}
\tilde{\b{U}}=\b{\dot{x}_0}-(\alpha \hat{x}-\b{\hat{s}}|_{s=0}).
\end{gather}

Since $\langle \alpha \hat{x} -\b{\hat{s}}|_{s=0}\rangle =0$ for periodic $\b{X}(s)$, this wave-frame velocity is equivalent in mean to the head velocity, $\langle \tilde{U} \rangle=\langle \dot{x}_0 \rangle=U$. The velocity of a point on the body in this special case may then be written as $\b{u}=(\tilde{U}+\alpha)\hat{x}-\b{\hat{s}}$, and Eqns.~\ref{Eqn:ZeroForceTorquea} reduce to
\begin{gather}
\tilde{U}=-\frac{\alpha(1-r_k)(1-\beta)}{1-(1-r_k)\beta},\label{x0defs}\\
\alpha=\int_0^1X_{s}\,ds,\,\,\,\,\beta=\int_0^1(X_{s})^2\,ds,\label{abdefs}\\
\dot{z}_0=-\hat{z}\cdot \b{\hat{s}}|_{s=0},\,\,\,\,\theta(t)=0.\label{z0defs}
\end{gather}
The rate of mechanical work is given by 
\begin{gather}
\Phi=(\tilde{U}+\alpha)^2(1-(1-r_k)\beta)+r_k(1-2\alpha(\tilde{U}+\alpha)).
\end{gather}
Defining the slope function $g(s)=Z_s(s)$, and setting the variational derivative of the efficiency to zero ($\delta \eta = 0$) the following algebraic equation is generated for $g(s)$:
\begin{equation}
g(s)^2=1-\frac{(1-\beta)^2(1-(1-r_k) \beta)^4}{r_k^2 \alpha^2(2-(1+r_k)\alpha^2-(1-r_k) (2-\alpha^2) \beta)^2}\,\,\,\,\,\mbox{(constant)}.
\label{Eqn:galgebraic}
\end{equation}
Since the absolute slope $|g(s)|$ is constant we may compute simply the constants $\beta=1-g^2=\alpha^2$. These relationships then yield $\beta=1/(1+\sqrt{r_k})$, and
\begin{gather}
g(s)=\pm\sqrt{\frac{\sqrt{r_k}}{1+\sqrt{r_k}}}\,\,\cdot
\end{gather}
This is the result of Lighthill \cite{Lighthill75}. For $r_k=1/2$, the physical slope of the sawtoothed waveform is $\sin^{-1}(|g(s)|)= 40.06^\circ$. The associated efficiency is $\eta=(1-\sqrt{r_k})^2=0.0858$, and the swimming velocity is $\tilde{U}=-(1-\sqrt{r_k})/(1+\sqrt{r_k})=-.224$, opposite the direction of the travelling wave. Generally speaking, a body shape which alternates between the positive and negative slopes $\pm |g(s)|$ at an arbitrary number of points yields the same swimming velocity and efficiency. The structure even admits pathologically discontinuous $g(s)$, up to the point at which the resistive force theory becomes invalid. This complication is removed in the case of infinite-length by simply requiring that a fundamental periodic mode is exhibited over a unit wavelength.

If the bending cost is included ($A_B>0$) but the sliding cost is ignored ($A_S=0$), after some algebra we find that the variational problem leads to the following integro-differential equation for the slope $g(s)$:
\begin{multline}
A_B \left(g_{ss}+\frac{gg_s^2}{1-g^2}\right)-\left(\frac{(1-A_B) r_k+(A_B/\gamma_{_B}) \mathcal{P}_{Bending}}{\alpha}\right)g\sqrt{1-g^2}\\
+ (1-A_B) r_k^2\left(\frac{\left(2-\alpha^2\right) (1-(1-r_k) \beta )-r_k \alpha^2}{(1-(1-r_k)\beta)^2(1-\beta)}\right)g(1-g^2)\\
+2\,r_k\,(A_B/\gamma_B) \left(\frac{\mathcal{P}_{Bending}  }{(1-(1-r_k) \beta )(1-\beta)}\right)g(1-g^2)=0.
\label{Eqn: ABgss}
\end{multline}
The $A_B\rightarrow 0$ limit is readily seen to be a singular one, which is expected due to the lack of regularity in the sawtoothed solution. A numerical optimization for the case of general $A_B>0$ is presented in a later section. However, rewriting Eqn.~\ref{Eqn: ABgss} for $A_B>0$ with its highest derivative alone on the left hand side makes clear that further differentiations introduce no irregularities for $|g(s)|<1$, so that such solutions have $g\in C^{\infty}$.

\subsection{Numerical optimization}
Having already assumed the fluid is modeled by the Stokes equations and since length scales out of the efficiency measure completely for a body of infinite length, there is a scale invariance in the determination of the optimal waveform. We account for this invariance in the infinite-length case by assuming that the waveform cannot be decomposed into smaller periodic forms. In other words, we scale the optimization so that the fundamental shape is expressed exactly once in the spatial period $s\in[0,1]$. This is achieved by considering the following basis for the tangent angle $\psi(s)$:
\begin{gather}
\psi(s)=\tan^{-1}\left(\frac{Z_s(s)}{X_s(s)}\right)=\sum_{n=1}^\infty a_n\cos (2\pi n s),
\end{gather}
subject to the constraint that $Z(s+1)=Z(s)$. We make a simplifying assumption that $\psi\in[-\pi,\pi]$. The solutions found without enforcing this constraint are consistent with this assumption, but the numerical search procedure can become unstable or slow in some cases without its application. The phase of the waveform is irrelevant for bodies of infinite-length. Given the tangent angle at each station $s$, we recover the flagellar shape by integration,
\begin{gather}
g(s)=Z_s(s)=\sin(\psi(s)),\,\,\,\,X_s=\cos(\psi(s)).
\end{gather}

Numerical optimization is performed using a SQP, Quasi-Newton, line-search method in MATLAB's optimization toolbox. The tangent angle is written as a finite sum over the first $n^*$ Fourier modes, and the corresponding coefficients $a_n$ are determined so that the efficiency $\eta$ is maximized. The optimization routine runs until the line search detects a local solution gradient with a relative error tolerance of $10^{-14}$.

The swimming velocity is determined by discretizing the slender filament in the arc-length $s$ by $M$ uniformly distributed points and solving the linear system, Eqn.~\ref{Eqn:ZeroForceTorquea}. Since the body is an infinitely long traveling wave, the velocity and rate of mechanical work are constant in time, and hence are determined numerically at $t=0$ alone. Hence, each iteration of the optimization routine requires the creation and inversion of only one algebraic equation for $\dot{x}_0(t)$, since $z_0(t)=\theta(t)=0$ in the infinite-length case. Generally, the values ($M=4000$, $n^*=80$) are sufficient so that further resolution has a negligible effect on the solution.

The constants ($\alpha$, $\beta$) as well as the rates of work $(\Phi,\mathcal{P}_{Bending})$ are determined by quadrature in $s$. Given that their integrated arguments are periodic on $s\in[0,1]$, a simple trapezoidal rule yields spectral accuracy. The optimization routine was seeded with a variety of initial flagellar shapes to increase the probability that a global maximum of efficiency was achieved. However, given enough spatial resolution the solutions found for the infinite-length case did not vary, regardless of the initial guess. In addition, the solutions so found have been verified by insertion into Eqn.~\ref{Eqn: ABgss}.

\subsection{Finite bending costs: numerical results}

\begin{figure}[ht]
\begin{center}
\includegraphics[width=4.5in]{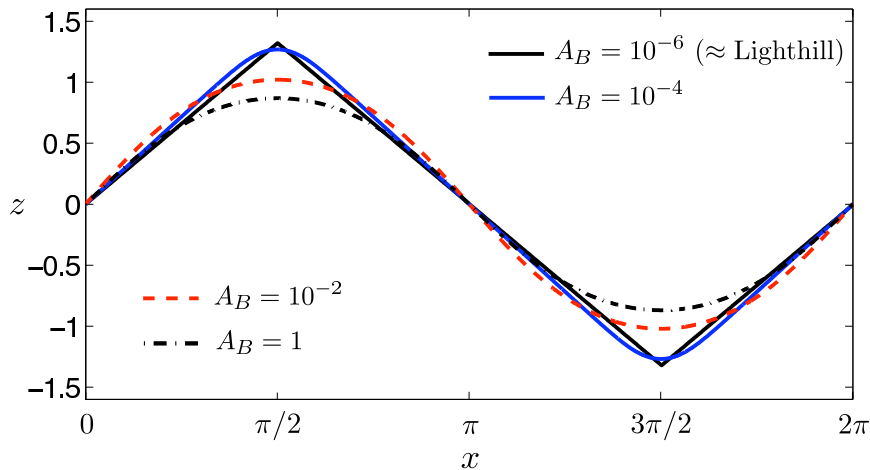}
\caption{(color online) Optimal waveforms for an infinite flagella with various bending costs ($A_B\neq 0$), in the case of no sliding cost ($A_S=0$). For a small bending cost ($A_B\ll 1$) the optimal waveform is very nearly the analytically derived sawtooth function. As the bending cost increases, the shape undergoes its most dramatic change for $A_B\sim 10^{-2}$, and settles to very nearly a sinusoid for $A_B=1$.}
\label{Figure5}
\end{center}
\end{figure}

We now present the optimal shapes of infinite-length bodies with the inclusion of the bending cost. Figure~\ref{Figure5} shows the optimal waveforms for a sequence of bending parameters, $A_B$, with no sliding cost ($A_S=0$). In order to best compare the shapes, the optimal waveforms are rescaled to the same physical wavelength for presentation.

\begin{figure}[ht]
\includegraphics[width=6.3in]{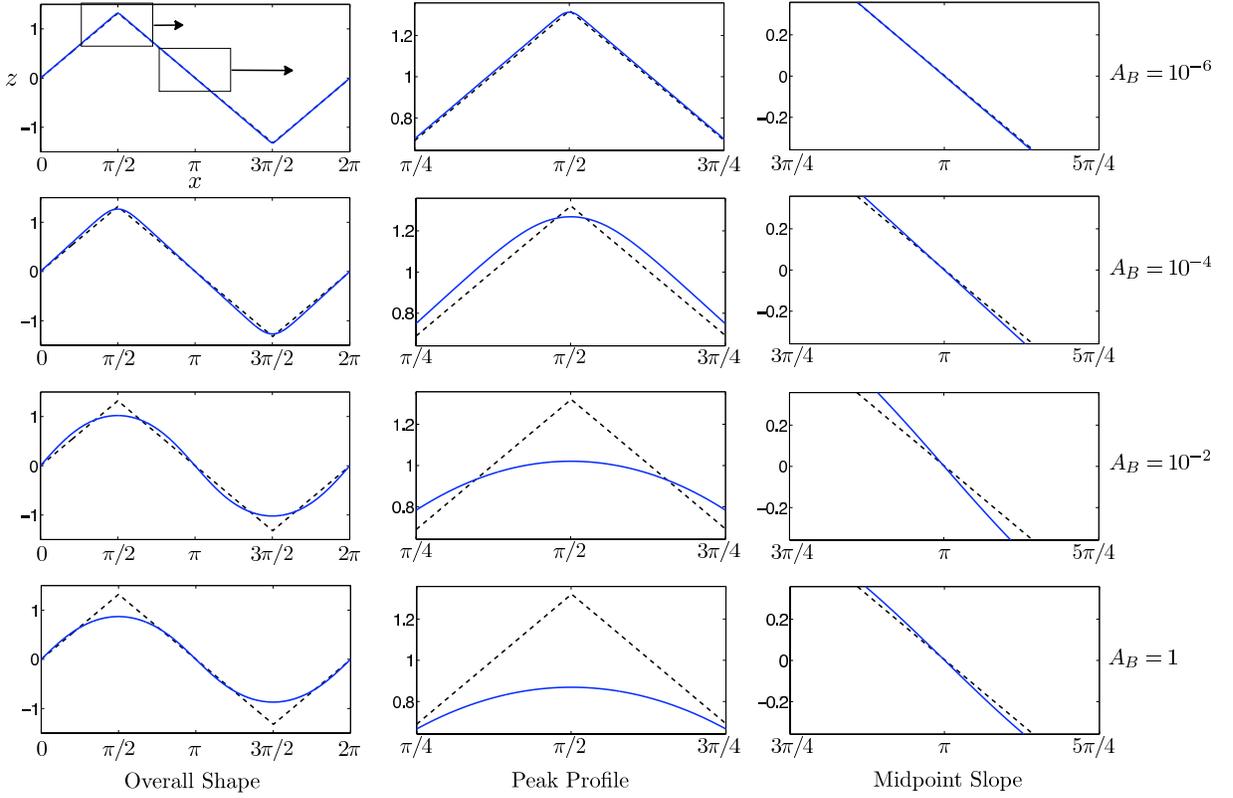}
\caption{(color online) A closer look at the
optimal waveform with the  inclusion of bending ($A_B\neq 0$) but no sliding costs ($A_S = 0$), and the  departure from the Lighthill result (dashed lines).}
\label{Figure6}
\end{figure}

For $A_B=10^{-6}$ the optimal waveform is very nearly the analytically derived sawtooth function. As the bending cost increases the shape undergoes its most dramatic change near $A_B\sim 10^{-2}$, and settles to very nearly a sinusoid for $A_B=1$; specifically, to the $Z(s)\approx .1159 \sin(2\pi s)-.0017 \sin(6\pi s)$, or $Z(x)\approx 0.1208\sin(2\pi x/\alpha)+0.0033\sin(6\pi x/\alpha)$, with $\alpha=0.85096$. The inclusion of bending costs as we propose in this paper therefore effectively regularizes the non-smoothness of Lighthill's solution.

\begin{figure}[ht]
\includegraphics[width=5.4in]{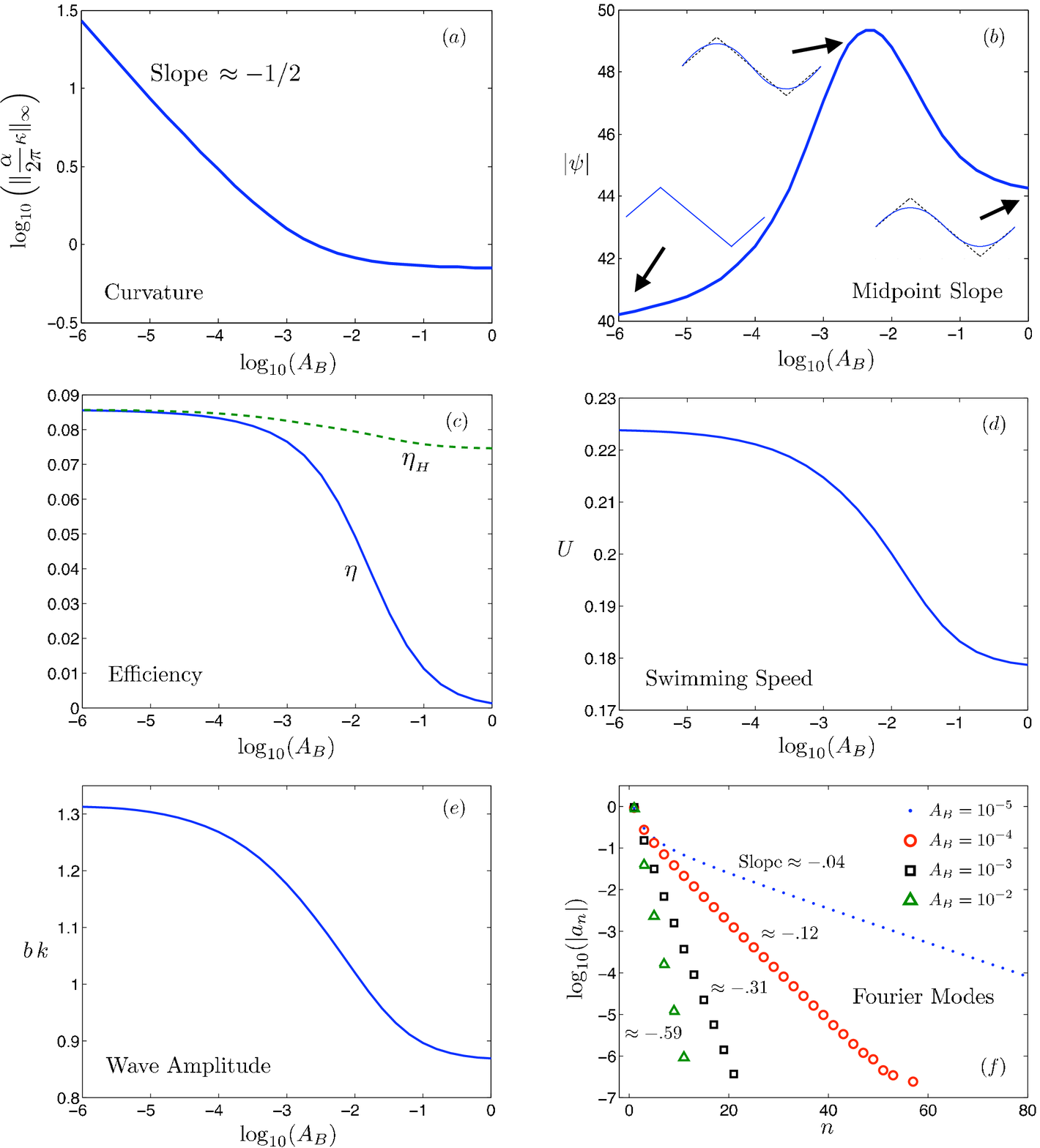}
\caption{(color online) Properties of the optimal waveform for an infinite flagellum as the bending parameter $A_B$ is varied from 0 to 1, with no sliding cost ($A_S=0$). 
(a) Normalized maximum curvature. 
(b) Absolute midpoint slope (in degrees); the behavior here is illustrated in the third column of Fig.~\ref{Figure6}. 
(c) Swimming efficiency $\eta$ and  hydrodynamic efficiency $\eta_H$.
(d) Swimming velocity $U$. 
(e) Normalized waveform amplitude $b\,k$, with $b=2\pi\|Z\|_\infty$ and $k=1/\alpha$ the wave-number. 
(f) Odd-numbered Fourier coefficients $|a_n|$ are shown to decay rapidly on a semi-logarithmic scale, indicating the solutions to be infinitely smooth. (Even-numbered modes are zero to working precision).}
\label{Figure7}
\end{figure}

In Figs.~\ref{Figure6} and \ref{Figure7}, we provide a closer inspection of the optimal shape and its properties for the same range of bending parameters $A_B$.  First, and as expected, with the additional cost of bending (increasing $A_B$), the curvature at the apex decreases from its infinite value in the Lighthill solution. The wavelength-normalized curvature at the apex is further plotted in Fig.~\ref{Figure7}a. It is not surprising to recover an apex curvature $\kappa\approx g_s \sim A_B^{-1/2}$ due to the form of Eqn.~\ref{Eqn: ABgss}. For $A_B\ll 1$, Eqn.~\ref{Eqn: ABgss} is dominated by the algebraic expression of Eqn.~\ref{Eqn:galgebraic} outside a boundary layer region where $A_B\, g_{ss} = O(1)$. The terms  in this equation are of like order in a region of size $s\sim \sqrt{A_B}$ around the discontinuities in the Lighthill solution. 

The third column of Fig.~\ref{Figure6} displays the slope at the midpoint of the body, which we find is not monotonic in the bending parameter $A_B$. This angle (absolute value) is further plotted in Fig.~\ref{Figure7}b. The absolute slope increases from $40.06^\circ$ to approximately $49^\circ$ at $A_B\approx 10^{-2.5}$, then decreases to nearly $44^\circ$ when the bending costs are prohibitively expensive ($A_B\rightarrow 1$). 

The total and hydrodynamic efficiencies are displayed in Fig.~\ref{Figure7}c as a function of the bending parameter $A_B$. For $A_B>10^{-2}$, the efficiency decreases approximately like $\eta\sim A_B^{-1}$, which may be predicted given an inspection of the efficiency measure $\eta$. The limiting value of the total efficiency is $0.00130$. Given the relatively small global change in the optimal shape, the hydrodynamic efficiency does not decrease as drastically as the total efficiency with increasing bending costs. For $A_B\rightarrow 1$, the hydrodynamic efficiency approaches the limit $\eta_H =0.0746$, a decay of only $13\%$ from the optimal Lighthill solution in the case where bending is without cost. In this case, the design of an organism or manmade swimmer is far more sensitive to the energetic costs due to bending than to hydrodynamic costs.

The swimming speed, $U$, is shown in Fig.~\ref{Figure7}d as a function of the bending parameter $A_B$, which decreases from the analytical solution in the previous section with increasing bending costs. Even though the hydrodynamic efficiency only decays $13\%$ from the optimal sawtoothed solution, the swimming speed decreases by approximately $20\%$. Fig.~\ref{Figure7}e shows the wavelength-normalized amplitude, $b\,k$, where $b=2\pi \|Z\|_\infty$ and $k=1/\alpha$ is the wave-number. This amplitude decreases from approximately $1.31$ to $0.87$ as $A_B\rightarrow 1$. 

Finally, Fig.~\ref{Figure7}f shows the odd-numbered Fourier modes in of the optimal waveform for four different bending parameters $A_B$ on semi-log axes. We observe that the Fourier coefficients decay linearly on this scale, so that $a_n \sim 10^{-p(A_B) n}$ for large (odd-numbered) $n$ and constants $p(A_B)$ as indicated in the figure. This decay, faster than polynomial in $n$, indicates that the optimal solution has $g(s)\in C^{\infty}$, in agreement with the previous comment (even-numbered Fourier modes are zero to working precision).

\subsection{Sinusoidal waveforms}

\begin{figure}[ht]
\includegraphics[width=6in]{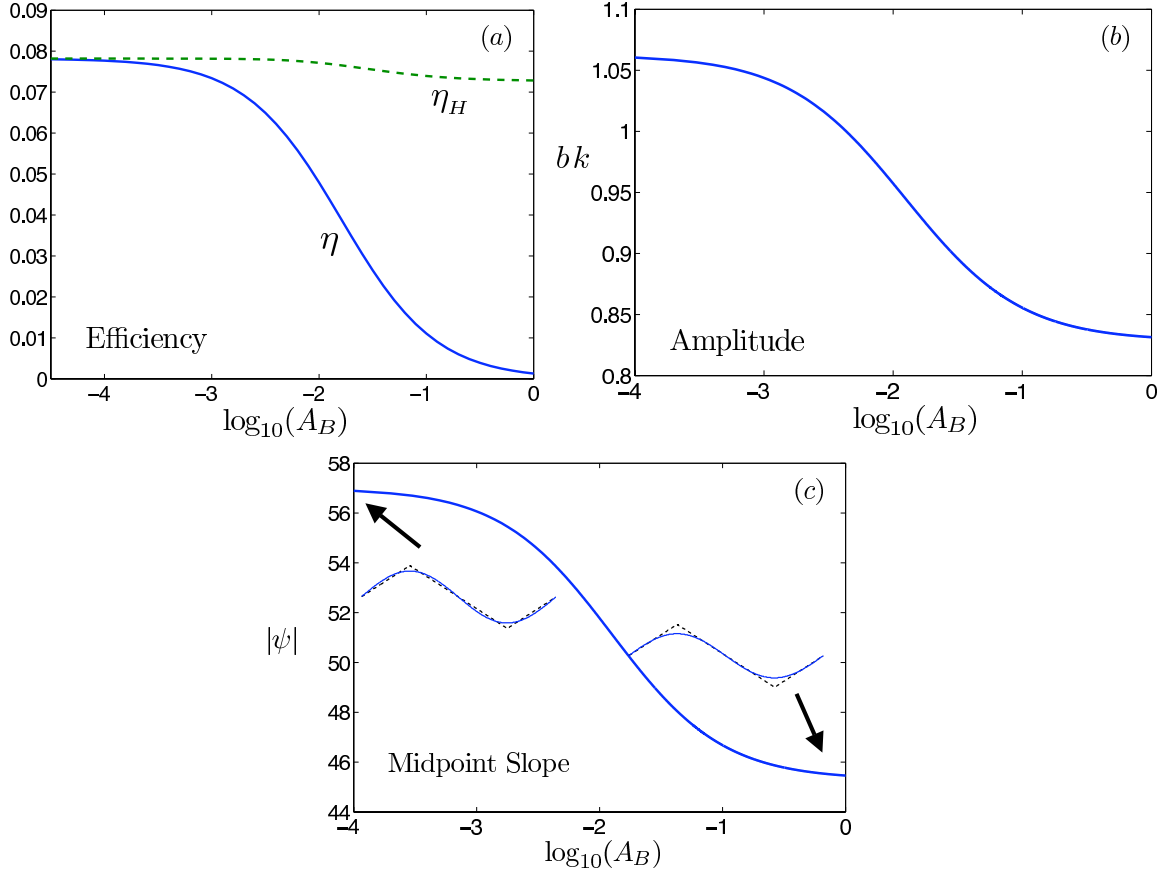}
\caption{(color online) Optimal sinusoidal waveforms of an infinite flagellum. 
 (a) The efficiency $\eta$ and hydrodynamic efficiency $\eta_{_H}$ of the optimal sinusoid are shown as functions of the bending parameter $A_B$. (b) Normalized amplitude $b\,k$. (c) Midpoint slope. }
\label{Figure8}
\end{figure}

The variational result given by Eqn.~\ref{Eqn: ABgss} does not lend itself to a straightforward long wavelength (small amplitude) asymptotic analysis; indeed, the solutions of interest require $g(s)= O(1)$ so that $g(s)\sim g(s)^3$. However, a linearization of Eqn.~\ref{Eqn: ABgss} balances the elasticity with the hydrodynamics: roughly, $g_{ss}+\omega g=0$. Hence, we expect periodic, near-sinusoidal solutions in general. This intuition is already corroborated by the near-sinusoidal solution for large bending parameters ($A_B\approx 1$), as previously discussed. For comparison, we compute the optimal amplitude of a sinusoidal waveform. Inserting the ansatz $Z(s)=\left(b/2\pi\right) \sin(2\pi s)$ into the efficiency measure, Eqn.~\ref{E:eta}, and using Eqns.~(\ref{x0defs}-\ref{z0defs}), yields
\begin{gather}
\alpha=\frac{2}{\pi}E(b^2),\,\,\,\,\,\beta=1-\frac{b^2}{2},\,\,\,\,\,\,\mathcal{P}_{Bending}=4\pi^2(1-\sqrt{1- b^2}),
\end{gather}
where $E(m)$ is the complete elliptic integral of the second kind, $(m\in[0,1])$. The slope amplitude $b$ that maximizes the efficiency is determined by setting $\partial_b \,\eta=0$. The expression for $\partial_b\, \eta$ is unwieldy so a rootfinding algorithm is used to locate the optimal amplitude as a function of the bending parameter and the resistance coefficient ratio $r_k$. The results are reported in Fig.~\ref{Figure8}. The optimal sinusoid when there are no bending costs ($A_B=0$) has an associated efficiency of $0.0782$, which is smaller than that of the optimal sawtoothed shape by only $9\%$. The total efficiency is seen to decrease with increasing bending costs to a limiting value of $0.00128$ when $A_B=1$, which is just barely smaller than the efficiency of the fully optimal solution determined by the numerical optimization ($\eta=0.00130$). Here again, the hydrodynamic efficiency does not decrease significantly as the bending becomes more expensive, since the optimal shape does not change dramatically. For $A_B \rightarrow 1$, we find $\eta_H=0.0728$.

The optimal amplitude decreases from a value of $b\,k = 1.06$ for $A_B=0$ to a limiting value of $b\,k =0.837$ when $A_B=1$. This matches very nearly the previous result in the numerical study ($b\,k=0.87$). Fig.~\ref{Figure8}c shows the midpoint slope angle, which is monotonic in the bending parameter $A_B$, and decreases from $57^\circ$ to $45^\circ$ as $A_B\rightarrow 1$, the latter result again nearly matching the result for the fully optimal shape.

\subsection{Finite internal sliding costs}

We now turn our attention to the consequences of finite internal sliding costs. Given the waveform periodicity the time-average of the sliding power may be simplified to an integration against the initial waveform. Recalling the travelling wave structure, and using that the tangent angle has zero mean, we have the expression
\begin{gather}
\mathcal{P}_{Sliding}=\gamma_{_S}\left(c_\Delta^2+ \int_0^1\int_0^1 \psi^2(s,t)\,ds\,dt\right).
\end{gather}
The inner integration may be written as independent of time by a simple manipulation,
\begin{gather}
\int_0^1 \int_0^1 \psi^2(s,t)\,ds\,dt=\int_0^1 \int_0^1 \psi^2(2\pi(s-t))\,ds\,dt=\int_0^1 \int_0^1 \psi^2(2\pi s)\,ds\,dt\\
=\int_0^1 \psi^2(2\pi s)\,ds=\int_0^1 \psi^2(s,0)\,ds.
\end{gather}
In other words, the sliding displacement travels with the waveform. In terms of the slope function $g(s)$,
\begin{gather}
\mathcal{P}_{Sliding}=\gamma_{_S}\left(c_\Delta^2+ \int_0^1 [\sin^{-1}(g(s))]^2\,ds\right).
\end{gather}

We are now prepared to consider the variational derivative of the efficiency when the sliding cost is included ($A_S>0$) but the bending cost is ignored ($A_B=0$). The result is once again an algebraic relation for the optimal shape:

\begin{equation}
c_1\, g(s)+c_2\, g(s)\sqrt{1-g^2(s)}+c_3\,\sin^{-1}(g(s))=0,
\label{Eq:Sliding}
\end{equation}
where we have defined
\begin{align}
c_1=&(1-\beta ) (1-(1-r_k) \beta )^2 ((1-A_S) r_k+(A_S/\gamma_{_S}) \mathcal{P}_{Sliding} ),\\
c_2=&r_k \alpha  \Big\{(1-A_S) r_k \left(\left(\alpha ^2-2\right)(1-(1-r_k)\beta )+r_k \alpha ^2 \right)\\
&\,\,\,\,\,\,\,\,\,\,\,\,\,\,\,\,\,\,\,\,\,\,\,\,\,\,\,\,\,\,-2 (A_S/\gamma_{_S}) (1-(1-r_k) \beta ) \mathcal{P}_{Sliding} \Big\},\\
c_3=&A_S \alpha  (1-\beta ) (1-(1-r_k) \beta )^2.
\end{align}

The form of Eqn.~\ref{Eq:Sliding} indicates that $|g(s)|=g_0$ (constant), so that once again we recover a solution which is not smooth at points where the slope switches sign. The inclusion of internal sliding costs does not, therefore, regularize the optimal swimming shape. For a sawtooth function we must have $\beta=\alpha^2=\sqrt{1-g_0^2}$. Inserting these values, the resulting expression is not amenable to analytical solution, but we apply a simple root-finding algorithm to determine the slope $g_0$ that satisfies Eqn.~\ref{Eq:Sliding}. 

\begin{figure}[ht]
\includegraphics[width=6in]{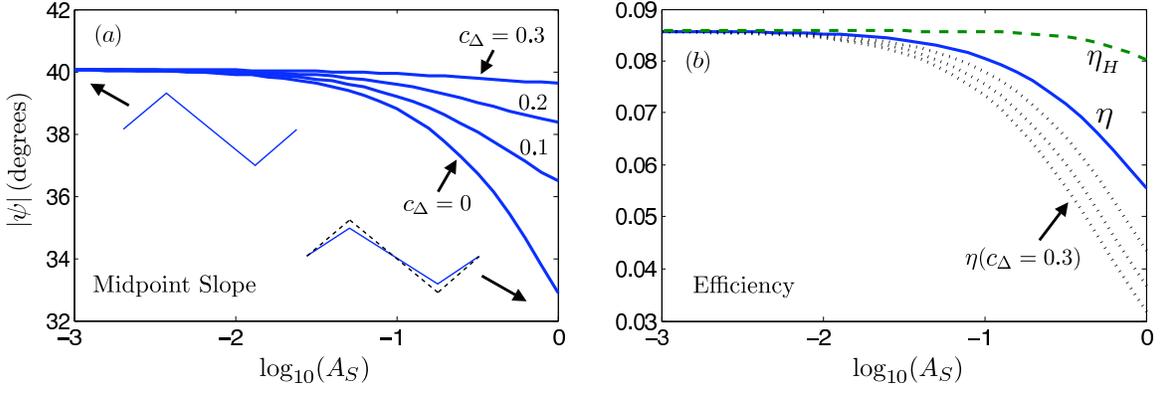}
\caption{The optimal infinite-length flagellar waveform with no bending cost ($A_B=0$) but with non-zero cost of elastic sliding ($A_S\neq 0$).
(a) As the sliding cost becomes more dominant (increasing $A_S$) the slope of the resulting sawtoothed waveform decreases from $40.06^\circ$ to a limiting value of approximately $33^\circ$ for zero base sliding ($c_\Delta=0$). If the base sliding is larger the sliding cost is relatively less dependent upon the waveform, and the optimal solution does not change as significantly with increasing $A_S$. (b) The total and hydrodynamic efficiencies for $c_\Delta=0$ are labelled. The efficiency decays significantly with increasing $A_S$ from the Lighthill value $\eta=0.0858$, but the hydrodynamic efficiency only decreases to a limit $0.0803$. Total efficiencies corresponding to the $c_\Delta$ constants in (a) are included as dotted lines.}
\label{Figure9}
\end{figure}

The results are reported in Figs.~\ref{Figure9}, where we show the resultant slope angle $\psi=\sin^{-1}(|g_0|)$ as well as the total and hydrodynamic efficiencies for a selection of constants $c_\Delta$. As $A_S \rightarrow 0$, the Lighthill solution is recovered in all cases: $\psi\approx 40^\circ$ and $\eta= 0.0858$. As the sliding cost becomes more dominant we find that the total efficiency drops significantly, and the optimal shape is a sawtooth function with a smaller and smaller amplitude. The slope angle $\psi$ decreases to nearly $33^\circ$ for $c_\Delta=0$ when the sliding costs vastly outweigh the hydrodynamic costs ($A_S\rightarrow 1$). For non-zero base sliding ($c_\Delta>0$) the sliding cost is relatively less dependent upon the waveform, and the optimal solution does not change as significantly with increasing $A_S$.

The total efficiency decays significantly with increasing $A_S$ from the Lighthill value $\eta=0.0858$ for $c_\Delta=0$, but the hydrodynamic efficiency only decreases to a limit $0.0803$. Since the optimal shape for very large sliding costs is similar to that when there are no sliding costs, it is not surprising to find that the hydrodynamic efficiency only begins to noticeably decay for $A_S > 10^{-1}$, and even then only to a limit $0.0803$. If the base sliding is larger, however, the waveform retains a nearly $40^\circ$ slope, and the hydrodynamic efficiency stays nearly constant for all values of $A_S$. As $A_S\rightarrow 1$ the efficiency scales logarithmically in $A_S$. 

\begin{figure}[ht]
\includegraphics[width=6in]{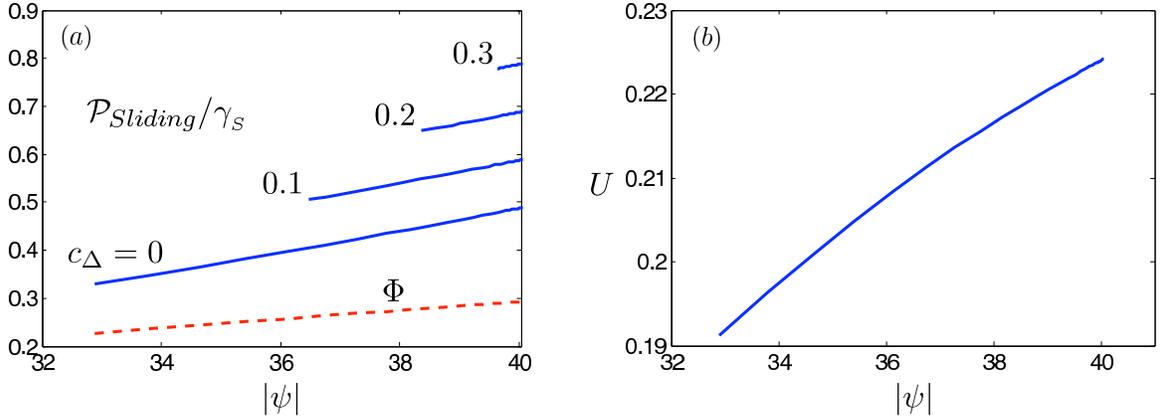}
\caption{(a) The sliding power as a function of the slope angle $|\psi|$ (degrees) for a selection of base sliding values $c_\Delta$, along with the rate of hydrodynamic work $\Phi$. (b) The swimming velocity as a function of the slope angle, limiting to the Lighthill limit as $A_S\rightarrow 0$ ($|\psi|\approx 40$).}
\label{Figure10}
\end{figure}

Figure~\ref{Figure10}a shows the sliding power $\mathcal{P}_{Sliding}/\gamma_{_S}$ for the same base sliding values shown in Figs.~\ref{Figure9} as functions of the slope angle, along with the rate of hydrodynamic work. For large sliding parameters $A_S\rightarrow 1$, the sliding power and rate of hydrodynamic work both decrease, but at the expense of a decrease in the swimming velocity. The decreasing swimming velocity as a function of the slope angle is shown in Fig.~\ref{Figure10}b.

\section{Bodies of Finite Length}\label{finite}

The passage of periodic waveforms down along a flagellum of finite length introduces new degrees of freedom, namely vertical net motions, and rotations. The introduction of rotation can break time-reversal symmetry for bodies of non-half-integer numbers of wavelengths. In order to generalize the waveform for finite-length flagella, we include a wavelength parameter in the specification of the shape. Specifically, for finite-size swimmers, we optimize over the first $n^*$ Fourier modes describing the tangent angle:
\begin{gather}
\psi(s)=\sum_{n=1}^{n^*} a_n\cos \left( 2\pi n\,k\, s \right)+b_n\sin \left(2\pi n\,k\,s \right),
\end{gather}
where the wave-number $k$ is to be determined as part of the optimization. Time is discretized into $T_M$ uniformly distributed points on the domain $t\in[0,1/k]$, and the body velocities $\b{\dot{x}_0}(t)$ and $\dot{\theta}(t)$ are determined at each time-step by inverting the three by three system, Eqns.~\ref{Eqn:ZeroForceTorquea}. The number of time-steps must be sufficient to capture the activity of the highest Fourier modes in the travelling wave solution. We also insert an important constraint for optimal 'swimming': we require $\theta(T)=\theta(0)$ so that the organism does not rotate in circles over many periods. However, as we will show, this does not remove the possibility of a slow vertical drift perpendicular to the initial body orientation. We generally use here $M=2400$ spatial gridpoints, $T_M=160$ time-steps, and $n^*=160$ Fourier modes. The solutions reported here were checked against simulations using more refined spatial and temporal discretizations and greater numbers of Fourier modes, when possible. The results did not vary significantly with further resolution. 

For bodies of finite length there is an important degeneracy in the model in the pure hydrodynamic consideration of  ($A_B=A_S=0$). Given a body of finite length, the optimal solution must in fact be Lighthill's sawtooth function with infinitesimally small amplitude and infinitely many wavelengths. In this limit, there is no rotation, and hence there are no hydrodynamical costs associated with rotational work done on the fluid. Any non-zero bending costs will regularize the geometry, and a competition between body rotations and the bending costs associated with the number of wavelengths expressed by the body will ensue. In contrast, as we have shown above, the inclusion of a sliding cost ($A_S>0$) does not regularize the optimal body shape. Hence, the optimal finite-length swimmer in the presence of a sliding cost must be the degenerate case of a body with infinitely-many wavelengths of infinitely small amplitude, with a slope as determined in the previous sections. We therefore consider below the influence of bending costs on the optimal finite-size flagellar waveform. The waveforms are not limited to any class of functions (other than periodic) and are determined by the numerical optimization.

\begin{figure}[ht]
\begin{center}
\includegraphics[width=3.5in]{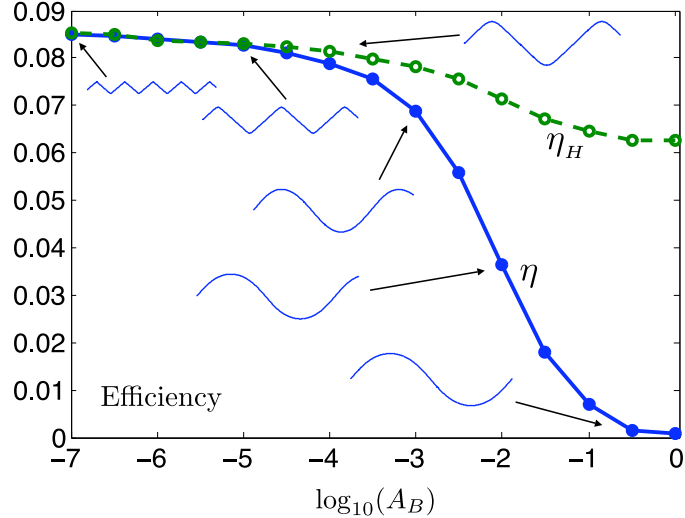}
\caption{(color online)  Swimming efficiencies for the optimal flagellum  of finite length as a function of the  bending cost $A_B$: total ($\eta$, solid line) and hydrodynamic ($\eta_H$, dashed line) efficiencies.}
\label{Figure11}
\end{center}
\end{figure}

\begin{figure}[ht]
\begin{center}
\includegraphics[width=6in]{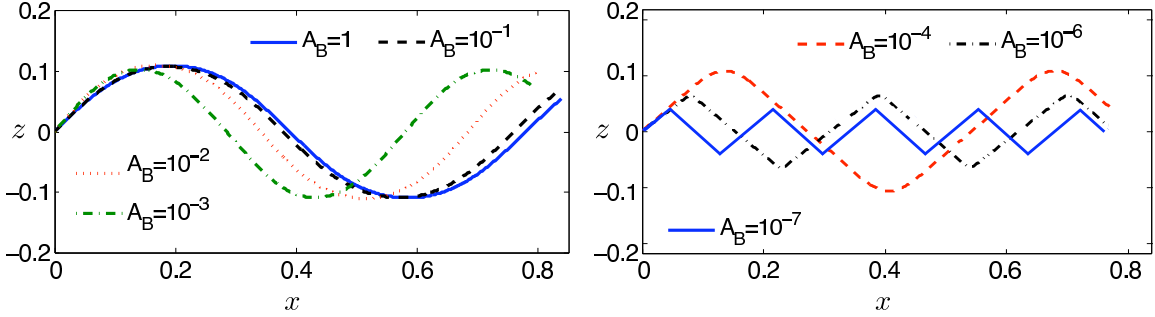}
\caption{(color online) Optimal finite-size flagellar waveforms for a selection of bending costs $A_B$. As bending becomes less costly (decreasing $A_B$), the optimal shape expresses larger wave-numbers and sharper profiles, with a bias towards half-integer wavelengths.}
\label{Figure12}
\end{center}
\end{figure}

Figure~\ref{Figure11} shows the total and hydrodynamic efficiencies associated with the optimal shapes for $A_B\in[10^{-7},1]$. The optimal shapes for a selection of bending costs are also included, and are presented for a more direct comparison in Figs.~\ref{Figure12}. When the bending cost is very large compared to the hydrodynamic cost ($A_B\sim 1$), the optimal shape in an approximate sinusoid and expresses just beyond a single wavelength, $k=1.08$. As the bending costs begin to decrease the optimal shape begins to express a slightly larger wave-number, $k=1.19$ for $A_B=10^{-2}$. With further decreases in the bending costs the optimal shape takes on a sharper profile, and approaches a half-integer wave-number, $k=1.42$ for $A_B=10^{-4}$. 

While the shape appears to change continuously for bending parameters in the range $A_B\in[10^{-4},1]$, we find a surprising transition between $A_B=10^{-4}$ and $A_B=10^{-4.5}$. While the efficiency appears to change continuously in this range of bending costs, the optimal shape jumps discontinuously to approximately the next half-integer wave-number. For $A_B>10^{-6}$ there is yet another discontinuous transition to the next half-integer wave-number, and the optimal shape becomes more and more like the infinite sawtoothed solution ($\eta=0.858$). We emphasize that this remarkable bias towards half-integer wave-numbers, which we observe over three full jumps, is an output of the optimization and not an assumed constraint.

We have observed that while the efficiency appears to be continuous through the jumps in wave-number, it is not smooth (not shown here). The total efficiency decreases to a limiting value of $\eta=7.68 \cdot 10^{-4}$ as the bending becomes exceedingly expensive ($A_B\rightarrow 1$). The hydrodynamic efficiency, on the other hand, decreases monotonically to a limiting value of $\eta_H=0.0606$ for $A_B\rightarrow 1$. The hydrodynamic efficiency decreases by $30\%$ in this limiting case, a more dramatic change than for the analogous body of infinite-length. Here the extra degrees of freedom, namely rotations and vertical drift, are more dependent upon body shape, and lead to larger variations in the dynamical work done to the fluid. Importantly however, this hydrodynamic efficiency of $\approx 6\%$ is still significantly above the efficiency of $\approx 1 \%$ typically displayed by biological cells which utilize planar waves \cite{Lighthill75,Childress81,tamPhD}.

\begin{figure}[ht]
\begin{center}
\includegraphics[width=5.8in]{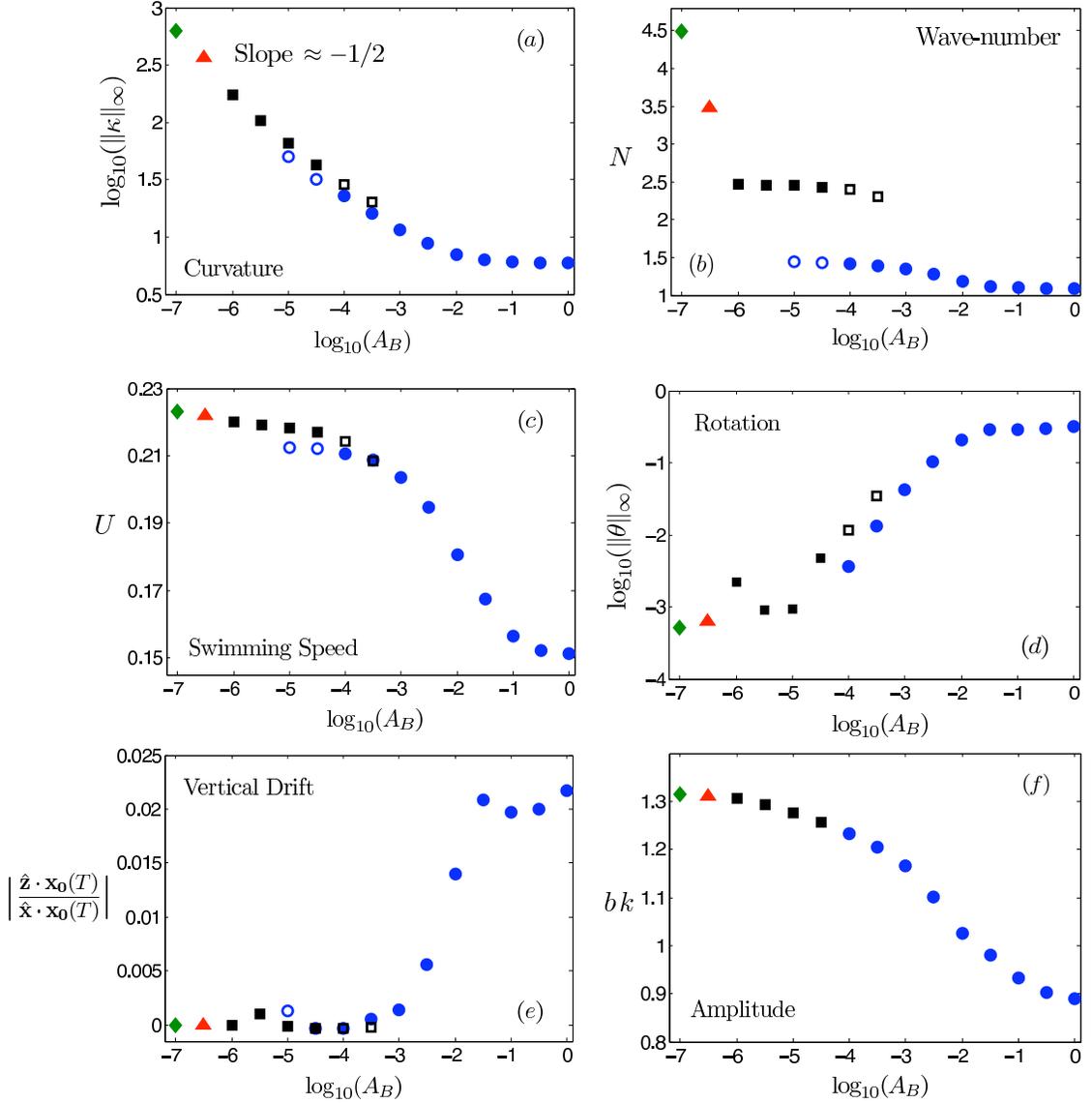}
\caption{(color online) Properties of the optimal finite-length flagellum as a function of the bending cost. The behavior limits to that of the degenerate, infinite-length solution as bending costs vanish ($A_B\rightarrow 0$).
(a) Maximum flagellum curvature.
 (b) Wave-number. 
 (c) Swimming speed. 
 (d) Maximum body rotation.
 (e) Vertical body drift.
 (f) Wavelength-normalized amplitude.
 }
\label{Figure13}
\end{center}
\end{figure}

Other properties of the optimal finite-length flagellum are shown in Figs.~\ref{Figure13}. Data corresponding to the optimal shape are shown as solid points, but we also include hollow points to indicate values for certain locally optimal solutions. Figure~\ref{Figure13}a shows the maximum curvature of the optimal shapes. Through the jump transition in wave-number there is a small jump in the maximum curvature, but the overall trend is preserved.  As in the infinite-length case, the maximum curvature of the finite-length body scales as approximately $\kappa \sim A_B^{-1/2}$ as $A_B\rightarrow 0$. However, particularly given the jumps in maximum curvature as the wave-number increases discontinuously the true asymptotic behavior as $A_B\rightarrow 0$ may not yet be well represented in this regime. As shown in Fig.~\ref{Figure13}b there is a distinct trend for decreasing bending costs towards half-integer wave-numbers, and there are also jumps to shapes of larger half-integer multiples at critical bending parameters. The left-right symmetry in shapes of half-integer wavelength serve to significantly decrease the body rotations throughout the motion. With smaller rotations, the body undulations can contribute more directly to lateral locomotion without performing much rotational work on the surrounding fluid. 

The swimming speed is shown in Fig.~\ref{Figure13}c. The locally optimal solutions with $k\approx 1.5$ (shown in circles) give way to the globally optimal solutions with $k\approx 2.5$ (squares) at approximately $A_B=10^{-4.5}$. The rate of increase in swimming speed for decreasing $A_B$ becomes more rapid with the larger wave-number solutions, so that the swimming speed appears continuous but non-smooth. The swimming speeds are for all $A_B$ smaller than the swimming speeds determined in the infinite-length consideration, as expected, since there are body rotations in the finite case which generally act to impede the lateral swimming motion. Figures~\ref{Figure13}d and e show the maximum rotation angle $||\theta(t)||_{\infty}$ and the vertical shift (or slope of the velocity vector) as functions of the bending cost. Given sinusoidal or sawtooth waveforms with integral numbers of wavelengths, it has been shown for small-amplitude motion that the rotation angle decreases with wave-number as $||\theta(t)||_{\infty}\sim 1/k^2$ \cite{pk74}. We have also found this scaling to hold for large amplitude waves (not shown). We observe non-monotonicity in the maximum rotation angle for $A_B\approx 10^{-1}$ and $A_B\approx 10^{-6}$. The vertical drift also exhibits non-monotonicity in the same regions. The behavior near $A_B=1$ is likely due to the transition from the extreme case of a single sinusoidal wavelength to the nearby (hydrodynamically preferred) half-integer wavelength. The small vertical drift of the swimming motion (also noted in \cite{dkb80}) is an effect which is third order in the wave amplitude for small amplitude waves, while the swimming velocity is second order in the amplitude; hence in small amplitude studies this drift is generally not observed. The drift decreases as the body takes on greater wave-numbers and undergoes smaller rotations. 

\begin{figure}[ht]
\begin{center}
\includegraphics[width=3.5in]{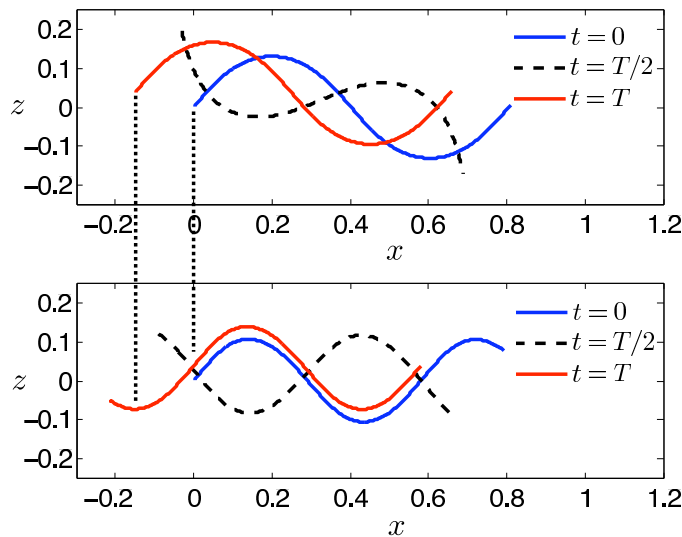}
\caption{(color online) Transition in the optimal waveform for a finite-size swimmer. 
(a) With $A_B=1$, the increased cost of bending leads to a smoother waveform, and in turn to significant rotation. $T=.994$ is the fundamental period of this first shape. (b) For $A_B=10^{-3}$, bending is not as energetically costly, and a higher spatial mode is observed to be optimal. This corresponds to a reduction in rotation, and thus a more efficient motion opposite the direction of the travelling wave. Nearly-half-integral wave numbers benefit from their approximate left-right symmetry, which significantly decreases rotations. $T=.994$, as in (a), for comparison.}
\label{Figure14}
\end{center}
\end{figure}

The hydrodynamic benefits of half-integer spatial modes is illustrated in Fig.~\ref{Figure14}. For $k\approx 1$ the body experiences a large rotation through the periodic motion, while the $k\approx 1.5$ mode for $A_B=10^{-3}$ shows damped rotations, and a more effective motion towards the left. The vertical drift is visible in the first case. 

Finally, Fig.~\ref{Figure13}f shows the wavelength-normalized amplitude of the optimal finite-length flagellum, $b\,k=2\pi \|Z\|_\infty\,k$. Perhaps surprisingly, even with the large rotations seen at small wave-numbers the optimal amplitude behaves very much like in the infinite-length case. For $A_B=1$ is very large the wave amplitude is $b\,k\approx 0.89$, just larger than the infinite-length result. As the bending costs decrease the optimal waveform approaches approximately the same limiting amplitude seen in Fig.~\ref{Figure7}e, and for $A_B=10^{-7}$ we find $b\,k=1.315$. Each jump to larger wave-numbers is accompanied by a jump in $\|Z\|_\infty$. Hence, the optimal waveform appears to degenerate towards Lighthill's infinite-length sawtooth solution in a self-similar fashion.

\section{Discussion}
\label{discussion} 

In this paper,  we have offered a physically-motivated derivation of the optimal flagellum shape.
We have considered the optimal shapes of periodic, planar flagellar waves of both infinite and finite length in a model which, in addition to hydrodynamic dissipation,  incorporates energetic costs of internal bending, sliding, and fluid dissipation. For bodies of infinite-length, we have shown that the inclusion of a bending cost (or dissipation due to the presence of an internal fluid) regularizes the classical Lighthill sawtooth solution, and that the optimal waveform becomes very nearly (but not quite) a sinusoid. The inclusion of a sliding cost has been shown to decrease the amplitude of the optimal waveform, but the optimal shape is still a sawtooth with a jump in the slope at a finite number of points. For bodies of finite-length, we have shown that a degenerate solution, in which the body takes on infinitely many small amplitude waves, is regularized by the addition of any bending cost (or internal fluid dissipation cost). With the exception of the case in which the bending is exceedingly expensive, the optimal shape has been shown to express approximately half-integer number of wavelengths, with a shape tending in a self-similar manner towards that of the infinite-length sawtoothed shape. This surprising result underlines the importance of minimizing the rotational work done on the surrounding fluid during lateral swimming. In addition, for both the infinite- and finite-length cases, we have shown that the change in the hydrodynamic efficiency is relatively small, and remains well above the hydrodynamic efficiency of typical biological cells.

The model presented here uses some simplifying assumptions, and leaves a number of open questions. First, the hydrodynamic description could be improved upon by the inclusion of non-local effects, for example using  slender body theory, or a more complete three-dimensional method for thicker organisms such as nematodes. The work of Tam \cite{tamPhD} appears to indicate that the sawtooth form may not be regularized by the non-local fluid interactions in the limit of zero bending costs, but that the number of expressed wavelengths may be decreased. Another exclusion in the work presented here is the possible presence of a head. Although most sperm cells have relatively small cell bodies (such as human spermatozoa), they can be large for some microorganisms and generally act to damp rotations imposed by the flagellar beating. In addition, the expression we used  for the bending energy becomes invalid when the radius of curvature approaches the body radius. The formation of material or structural singularities has been considered by other authors, and this can also provide a barrier to the degeneracy mentioned above \cite{gd07}. 

\begin{figure}[ht]
\begin{center}
\includegraphics[width=3in]{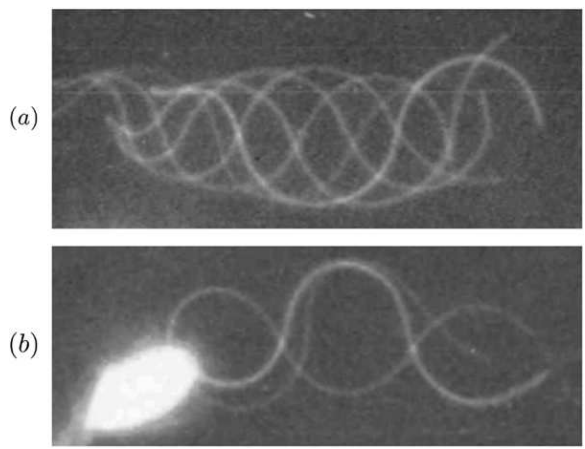}
\caption{Marine invertebrates spermatozoa. (a): Superimposed images of the headless spermatozoon of \textit{Lytechinus}. (b): Spermatozoon of \textit{Chaetopterus} exhibits non-integral spatial wave-numbers.
Reproduced with permission by the Journal of Experimental Biology, Ref.~\cite{Brokaw65}.}
\label{Figure15}
\end{center}
\end{figure}

In our opinion, the two most important implications of our study for the biophysics of swimming cells are the following. First, we have shown that a physically-motivated measure of internal elastic cost for the deformable flagellum regularizes the hydrodynamically-optimal solution of Lighthill, and that this is done with only a small loss in hydrodynamic efficiency. Second, our results show the emergence of small numbers of wavelengths in the optimal solution when bending is at all costly (see Fig.~\ref{Figure11}). This result, which is likely to remain robust to improvements on the modeling such as the one discussed above, is consistent with observations on the morphology of eukaryotic flagella, in particular for spermatozoa (see the review in Ref.~\cite{bw77}). As an example, we reproduce in Fig.~\ref{Figure15} the shapes of two marine invertebrates spermatozoa  (\textit{Lytechinus} and \textit{Chaetopterus}) from Ref.~\cite{Brokaw65}. In both cases, although the shapes are different from out optimal solutions, the presence of the half-integer wave-number morphology ($k\approx 1.5$) is apparent. Our work constitutes therefore an attempt at a physical rationalization of this observed feature of eukaryotic flagella. We also observe that our optimal solutions display hydrodynamic efficiencies which are significantly above those of biological swimming cells, which are typically in the 1\% range. Our solution could therefore also be considered as an appropriate `initial condition' for further (more directly biological) optimization, at the expense of hydrodynamic efficiency. Finally, we note that another (less common) means of eukaryotic propulsion involves the passage of periodic helical waves down along the length of a flagellum. In this case, in addition to the costs discussed here, there may also be costs due to a twisting of the material, and dynein motors have been observed in some cases to exert twisting moments on the axoneme \cite{hb85}. The optimal shape of a helical flagellum under these energetic constraints will be considered in a future work.

\section*{Acknowledgements}
We thank Daniel Tam for useful discussions and Charles Brokaw for the use of his images. We gratefully acknowledge the support of the National Science Foundation through the grants CTS-0624830 and CBET-0746285.

\bibliography{OptimalFlagellum}

\begin{thebibliography}{43}
\expandafter\ifx\csname natexlab\endcsname\relax\def\natexlab#1{#1}\fi
\expandafter\ifx\csname bibnamefont\endcsname\relax
  \def\bibnamefont#1{#1}\fi
\expandafter\ifx\csname bibfnamefont\endcsname\relax
  \def\bibfnamefont#1{#1}\fi
\expandafter\ifx\csname citenamefont\endcsname\relax
  \def\citenamefont#1{#1}\fi
\expandafter\ifx\csname url\endcsname\relax
  \def\url#1{\texttt{#1}}\fi
\expandafter\ifx\csname urlprefix\endcsname\relax\def\urlprefix{URL }\fi
\providecommand{\bibinfo}[2]{#2}
\providecommand{\eprint}[2][]{\url{#2}}

\bibitem[{\citenamefont{Lauga and Powers}(2009)}]{lp09}
\bibinfo{author}{\bibfnamefont{E.}~\bibnamefont{Lauga}} \bibnamefont{and}
  \bibinfo{author}{\bibfnamefont{T.}~\bibnamefont{Powers}},
  \bibinfo{journal}{Rep. Prog. Phys.} \textbf{\bibinfo{volume}{72}}
  (\bibinfo{year}{2009}).

\bibitem[{\citenamefont{Purcell}(1977)}]{Purcell77}
\bibinfo{author}{\bibfnamefont{E.~M.} \bibnamefont{Purcell}},
  \bibinfo{journal}{Am. J. Phys.} \textbf{\bibinfo{volume}{45}},
  \bibinfo{pages}{3} (\bibinfo{year}{1977}).

\bibitem[{\citenamefont{Lighthill}(1975)}]{Lighthill75}
\bibinfo{author}{\bibfnamefont{J.}~\bibnamefont{Lighthill}},
  \emph{\bibinfo{title}{Mathematical Biofluiddynamics}}
  (\bibinfo{publisher}{SIAM}, \bibinfo{address}{Philadelphia},
  \bibinfo{year}{1975}).

\bibitem[{\citenamefont{Childress}(1981)}]{Childress81}
\bibinfo{author}{\bibfnamefont{S.}~\bibnamefont{Childress}},
  \emph{\bibinfo{title}{Mechanics of {S}wimming and {F}lying}}
  (\bibinfo{publisher}{Cambridge University Press}, \bibinfo{address}{Cambridge
  U.K.}, \bibinfo{year}{1981}).

\bibitem[{\citenamefont{Macnab}(1976)}]{Macnab76}
\bibinfo{author}{\bibfnamefont{R.~M.} \bibnamefont{Macnab}},
  \bibinfo{journal}{J. Clin. Microbiol.} \textbf{\bibinfo{volume}{4}},
  \bibinfo{pages}{258} (\bibinfo{year}{1976}).

\bibitem[{\citenamefont{Block et~al.}(1991)\citenamefont{Block, Fahrner, and
  Berg}}]{bfb91}
\bibinfo{author}{\bibfnamefont{S.}~\bibnamefont{Block}},
  \bibinfo{author}{\bibfnamefont{K.}~\bibnamefont{Fahrner}}, \bibnamefont{and}
  \bibinfo{author}{\bibfnamefont{H.}~\bibnamefont{Berg}}, \bibinfo{journal}{J.
  Bacteriology} \textbf{\bibinfo{volume}{173}}, \bibinfo{pages}{933}
  (\bibinfo{year}{1991}).

\bibitem[{\citenamefont{Turner et~al.}(2000)\citenamefont{Turner, Ryu, and
  Berg}}]{twb00}
\bibinfo{author}{\bibfnamefont{L.}~\bibnamefont{Turner}},
  \bibinfo{author}{\bibfnamefont{W.~S.} \bibnamefont{Ryu}}, \bibnamefont{and}
  \bibinfo{author}{\bibfnamefont{H.~C.} \bibnamefont{Berg}},
  \bibinfo{journal}{J. Bacteriol.} \textbf{\bibinfo{volume}{182}},
  \bibinfo{pages}{2793} (\bibinfo{year}{2000}).

\bibitem[{\citenamefont{Taylor}(1951)}]{Taylor51}
\bibinfo{author}{\bibfnamefont{G.~I.} \bibnamefont{Taylor}},
  \bibinfo{journal}{Proc. Roy. Soc. A} \textbf{\bibinfo{volume}{209}},
  \bibinfo{pages}{447} (\bibinfo{year}{1951}).

\bibitem[{\citenamefont{Hancock}(1953)}]{Hancock53}
\bibinfo{author}{\bibfnamefont{G.~J.} \bibnamefont{Hancock}},
  \bibinfo{journal}{Proc. Roy. Soc. Lond. A} \textbf{\bibinfo{volume}{217}},
  \bibinfo{pages}{96} (\bibinfo{year}{1953}).

\bibitem[{\citenamefont{Gray}(1955)}]{Gray55}
\bibinfo{author}{\bibfnamefont{J.}~\bibnamefont{Gray}}, \bibinfo{journal}{J.
  Exp. Biol.} \textbf{\bibinfo{volume}{32}}, \bibinfo{pages}{775}
  (\bibinfo{year}{1955}).

\bibitem[{\citenamefont{Lighthill}(1976)}]{Lighthill76}
\bibinfo{author}{\bibfnamefont{J.}~\bibnamefont{Lighthill}},
  \bibinfo{journal}{SIAM Rev.} \textbf{\bibinfo{volume}{18}},
  \bibinfo{pages}{161} (\bibinfo{year}{1976}).

\bibitem[{\citenamefont{Batchelor}(1970)}]{Batchelor70}
\bibinfo{author}{\bibfnamefont{G.}~\bibnamefont{Batchelor}},
  \bibinfo{journal}{J. Fluid Mech.} \textbf{\bibinfo{volume}{44}},
  \bibinfo{pages}{419} (\bibinfo{year}{1970}).

\bibitem[{\citenamefont{Cox}(1970)}]{Cox70}
\bibinfo{author}{\bibfnamefont{R.~G.} \bibnamefont{Cox}}, \bibinfo{journal}{J.
  Fluid Mech.} \textbf{\bibinfo{volume}{44}}, \bibinfo{pages}{791}
  (\bibinfo{year}{1970}).

\bibitem[{\citenamefont{Keller and Rubinow}(1976)}]{kr76}
\bibinfo{author}{\bibfnamefont{J.~B.} \bibnamefont{Keller}} \bibnamefont{and}
  \bibinfo{author}{\bibfnamefont{S.~I.} \bibnamefont{Rubinow}},
  \bibinfo{journal}{Biophys J.} \textbf{\bibinfo{volume}{16}},
  \bibinfo{pages}{151Ð170} (\bibinfo{year}{1976}).

\bibitem[{\citenamefont{Johnson}(1980)}]{Johnson80}
\bibinfo{author}{\bibfnamefont{R.~E.} \bibnamefont{Johnson}},
  \bibinfo{journal}{J. Fluid Mech.} \textbf{\bibinfo{volume}{99}},
  \bibinfo{pages}{411} (\bibinfo{year}{1980}).

\bibitem[{\citenamefont{Machin}(1958)}]{Machin58}
\bibinfo{author}{\bibfnamefont{K.~E.} \bibnamefont{Machin}},
  \bibinfo{journal}{J. Exp. Biol} \textbf{\bibinfo{volume}{35}},
  \bibinfo{pages}{796} (\bibinfo{year}{1958}).

\bibitem[{\citenamefont{Higdon}(1979)}]{Higdon79}
\bibinfo{author}{\bibfnamefont{J.~J.~L.} \bibnamefont{Higdon}},
  \bibinfo{journal}{J. Fluid Mech.} \textbf{\bibinfo{volume}{90}},
  \bibinfo{pages}{685} (\bibinfo{year}{1979}).

\bibitem[{\citenamefont{Brokaw}(1965)}]{Brokaw65}
\bibinfo{author}{\bibfnamefont{C.}~\bibnamefont{Brokaw}}, \bibinfo{journal}{J.
  Exp. Biol.} \textbf{\bibinfo{volume}{43}}, \bibinfo{pages}{455}
  (\bibinfo{year}{1965}).

\bibitem[{\citenamefont{Brokaw}(1970)}]{Brokaw70}
\bibinfo{author}{\bibfnamefont{C.}~\bibnamefont{Brokaw}}, \bibinfo{journal}{J.
  Exp. Biol.} \textbf{\bibinfo{volume}{53}}, \bibinfo{pages}{445}
  (\bibinfo{year}{1970}).

\bibitem[{\citenamefont{Brokaw}(1972)}]{Brokaw72}
\bibinfo{author}{\bibfnamefont{C.}~\bibnamefont{Brokaw}},
  \bibinfo{journal}{Biophys. J.} \textbf{\bibinfo{volume}{12}},
  \bibinfo{pages}{564} (\bibinfo{year}{1972}).

\bibitem[{\citenamefont{Brennen and Winet}(1977)}]{bw77}
\bibinfo{author}{\bibfnamefont{C.}~\bibnamefont{Brennen}} \bibnamefont{and}
  \bibinfo{author}{\bibfnamefont{H.}~\bibnamefont{Winet}},
  \bibinfo{journal}{Ann. Rev. Fluid Mech.} \textbf{\bibinfo{volume}{9}},
  \bibinfo{pages}{339} (\bibinfo{year}{1977}).

\bibitem[{\citenamefont{Camalet and Julicher}(2000)}]{cj00}
\bibinfo{author}{\bibfnamefont{S.}~\bibnamefont{Camalet}} \bibnamefont{and}
  \bibinfo{author}{\bibfnamefont{F.}~\bibnamefont{Julicher}},
  \bibinfo{journal}{New J. Phys.} \textbf{\bibinfo{volume}{2}},
  \bibinfo{pages}{1} (\bibinfo{year}{2000}).

\bibitem[{\citenamefont{Reidel-Kruse et~al.}(2007)\citenamefont{Reidel-Kruse,
  Hilfinger, Howard, and Julicher}}]{rhhj07}
\bibinfo{author}{\bibfnamefont{I.}~\bibnamefont{Reidel-Kruse}},
  \bibinfo{author}{\bibfnamefont{A.}~\bibnamefont{Hilfinger}},
  \bibinfo{author}{\bibfnamefont{J.}~\bibnamefont{Howard}}, \bibnamefont{and}
  \bibinfo{author}{\bibfnamefont{F.}~\bibnamefont{Julicher}},
  \bibinfo{journal}{HFSP J.} \textbf{\bibinfo{volume}{1}}, \bibinfo{pages}{192}
  (\bibinfo{year}{2007}).

\bibitem[{\citenamefont{Chaudhury}(1979)}]{Chaudhury79}
\bibinfo{author}{\bibfnamefont{T.~K.} \bibnamefont{Chaudhury}},
  \bibinfo{journal}{J. Fluid Mech.} \textbf{\bibinfo{volume}{95}},
  \bibinfo{pages}{189} (\bibinfo{year}{1979}).

\bibitem[{\citenamefont{Sturges}(1981)}]{Sturges81}
\bibinfo{author}{\bibfnamefont{L.}~\bibnamefont{Sturges}}, \bibinfo{journal}{J.
  Non-Newt. Fluid Mech.} \textbf{\bibinfo{volume}{8}}, \bibinfo{pages}{357}
  (\bibinfo{year}{1981}).

\bibitem[{\citenamefont{Fulford et~al.}(1998)\citenamefont{Fulford, Katz, and
  Powell}}]{fkp98}
\bibinfo{author}{\bibfnamefont{G.~R.} \bibnamefont{Fulford}},
  \bibinfo{author}{\bibfnamefont{D.~F.} \bibnamefont{Katz}}, \bibnamefont{and}
  \bibinfo{author}{\bibfnamefont{R.~L.} \bibnamefont{Powell}},
  \bibinfo{journal}{Biorheol.} \textbf{\bibinfo{volume}{35}},
  \bibinfo{pages}{295} (\bibinfo{year}{1998}).

\bibitem[{\citenamefont{Lauga}(2007)}]{Lauga07}
\bibinfo{author}{\bibfnamefont{E.}~\bibnamefont{Lauga}},
  \bibinfo{journal}{Phys. Fluids} \textbf{\bibinfo{volume}{19}},
  \bibinfo{pages}{083104} (\bibinfo{year}{2007}).

\bibitem[{\citenamefont{Fu et~al.}(2007)\citenamefont{Fu, Powers, and
  Wolgemuth}}]{fpw07}
\bibinfo{author}{\bibfnamefont{H.~C.} \bibnamefont{Fu}},
  \bibinfo{author}{\bibfnamefont{T.~R.} \bibnamefont{Powers}},
  \bibnamefont{and} \bibinfo{author}{\bibfnamefont{H.~C.}
  \bibnamefont{Wolgemuth}}, \bibinfo{journal}{Phys. Rev. Lett.}
  \textbf{\bibinfo{volume}{99}}, \bibinfo{pages}{258101}
  (\bibinfo{year}{2007}).

\bibitem[{\citenamefont{Smith et~al.}(2009)\citenamefont{Smith, Gaffney,
  Gadelha, Kapur, and Kirkman-Brown}}]{sg09}
\bibinfo{author}{\bibfnamefont{D.}~\bibnamefont{Smith}},
  \bibinfo{author}{\bibfnamefont{E.}~\bibnamefont{Gaffney}},
  \bibinfo{author}{\bibfnamefont{H.}~\bibnamefont{Gadelha}},
  \bibinfo{author}{\bibfnamefont{N.}~\bibnamefont{Kapur}}, \bibnamefont{and}
  \bibinfo{author}{\bibfnamefont{J.}~\bibnamefont{Kirkman-Brown}},
  \bibinfo{journal}{Cell Motility and the Cytoskeleton}
  \textbf{\bibinfo{volume}{66}}, \bibinfo{pages}{220} (\bibinfo{year}{2009}).

\bibitem[{\citenamefont{Pironneau and Katz}(1974)}]{pk74}
\bibinfo{author}{\bibfnamefont{O.}~\bibnamefont{Pironneau}} \bibnamefont{and}
  \bibinfo{author}{\bibfnamefont{D.~F.} \bibnamefont{Katz}},
  \bibinfo{journal}{J. Fluid Mech.} \textbf{\bibinfo{volume}{66}},
  \bibinfo{pages}{391} (\bibinfo{year}{1974}).

\bibitem[{\citenamefont{Silvester and Holwill}(1972)}]{sh72}
\bibinfo{author}{\bibfnamefont{N.}~\bibnamefont{Silvester}} \bibnamefont{and}
  \bibinfo{author}{\bibfnamefont{M.}~\bibnamefont{Holwill}},
  \bibinfo{journal}{J. Theor. Biol.} \textbf{\bibinfo{volume}{35}},
  \bibinfo{pages}{505} (\bibinfo{year}{1972}).

\bibitem[{\citenamefont{Dresdner et~al.}(1980)\citenamefont{Dresdner, Katz, and
  Berger}}]{dkb80}
\bibinfo{author}{\bibfnamefont{R.~D.} \bibnamefont{Dresdner}},
  \bibinfo{author}{\bibfnamefont{D.~F.} \bibnamefont{Katz}}, \bibnamefont{and}
  \bibinfo{author}{\bibfnamefont{S.~A.} \bibnamefont{Berger}},
  \bibinfo{journal}{J. Fluid Mech.} \textbf{\bibinfo{volume}{97}},
  \bibinfo{pages}{591} (\bibinfo{year}{1980}).

\bibitem[{\citenamefont{Tam}(2008)}]{tamPhD}
\bibinfo{author}{\bibfnamefont{D.~S.-W.} \bibnamefont{Tam}}, Ph.D. thesis,
  \bibinfo{school}{Massachusetts Institute of Technology},
  \bibinfo{address}{Cambridge, MA} (\bibinfo{year}{2008}).

\bibitem[{\citenamefont{Gray and Hancock}(1955)}]{gh55}
\bibinfo{author}{\bibfnamefont{J.}~\bibnamefont{Gray}} \bibnamefont{and}
  \bibinfo{author}{\bibfnamefont{G.~J.} \bibnamefont{Hancock}},
  \bibinfo{journal}{J. Exp. Biol.} \textbf{\bibinfo{volume}{32}},
  \bibinfo{pages}{802} (\bibinfo{year}{1955}).

\bibitem[{\citenamefont{Johnson and Brokaw}(1979)}]{jb79}
\bibinfo{author}{\bibfnamefont{R.~E.} \bibnamefont{Johnson}} \bibnamefont{and}
  \bibinfo{author}{\bibfnamefont{C.~J.} \bibnamefont{Brokaw}},
  \bibinfo{journal}{Biophys. J.} \textbf{\bibinfo{volume}{25}},
  \bibinfo{pages}{113 } (\bibinfo{year}{1979}).

\bibitem[{\citenamefont{Pozrikidis}(1992)}]{Pozrikidis92}
\bibinfo{author}{\bibfnamefont{C.}~\bibnamefont{Pozrikidis}},
  \emph{\bibinfo{title}{Boundary Integral and Singularity Methods for
  Linearized Viscous Flow}} (\bibinfo{publisher}{Cambridge University Press},
  \bibinfo{address}{Cambridge, UK}, \bibinfo{year}{1992}).

\bibitem[{\citenamefont{Brokaw}(1989)}]{Brokaw89}
\bibinfo{author}{\bibfnamefont{C.}~\bibnamefont{Brokaw}},
  \bibinfo{journal}{Science} \textbf{\bibinfo{volume}{243}},
  \bibinfo{pages}{1593} (\bibinfo{year}{1989}).

\bibitem[{\citenamefont{Landau and Lifshitz}(1986)}]{ll86}
\bibinfo{author}{\bibfnamefont{L.}~\bibnamefont{Landau}} \bibnamefont{and}
  \bibinfo{author}{\bibfnamefont{E.}~\bibnamefont{Lifshitz}},
  \emph{\bibinfo{title}{Theory of Elasticity, 3rd ed.}}
  (\bibinfo{publisher}{Pergamon Press}, \bibinfo{address}{Oxford},
  \bibinfo{year}{1986}).

\bibitem[{\citenamefont{Ghatak and Das}(2007)}]{gd07}
\bibinfo{author}{\bibfnamefont{A.}~\bibnamefont{Ghatak}} \bibnamefont{and}
  \bibinfo{author}{\bibfnamefont{A.}~\bibnamefont{Das}},
  \bibinfo{journal}{Phys. Rev. Lett.} \textbf{\bibinfo{volume}{99}},
  \bibinfo{pages}{076101} (\bibinfo{year}{2007}).

\bibitem[{\citenamefont{Brokaw}(1994)}]{Brokaw94}
\bibinfo{author}{\bibfnamefont{C.}~\bibnamefont{Brokaw}},
  \bibinfo{journal}{Cell Motil. Cytoskeleton} \textbf{\bibinfo{volume}{28}},
  \bibinfo{pages}{199} (\bibinfo{year}{1994}).

\bibitem[{\citenamefont{Hilfinger}(2005)}]{HilfingerPhD}
\bibinfo{author}{\bibfnamefont{A.}~\bibnamefont{Hilfinger}}, Ph.D. thesis,
  \bibinfo{school}{Dresden University of Technology},
  \bibinfo{address}{Dresden, Germany} (\bibinfo{year}{2005}).

\bibitem[{\citenamefont{Brokaw}(1971)}]{Brokaw71Sliding}
\bibinfo{author}{\bibfnamefont{C.}~\bibnamefont{Brokaw}}, \bibinfo{journal}{J.
  Exp. Biol.} \textbf{\bibinfo{volume}{55}}, \bibinfo{pages}{289}
  (\bibinfo{year}{1971}).

\bibitem[{\citenamefont{Hines and Blum}(1985)}]{hb85}
\bibinfo{author}{\bibfnamefont{M.}~\bibnamefont{Hines}} \bibnamefont{and}
  \bibinfo{author}{\bibfnamefont{J.}~\bibnamefont{Blum}},
  \bibinfo{journal}{Biophys. J.} \textbf{\bibinfo{volume}{47}},
  \bibinfo{pages}{705} (\bibinfo{year}{1985}).

\end{thebibliography}

\end{document}